\documentclass[aps,twocolumn,showpacs,showkeys]{revtex4}
\newcommand{\bfm}[1]{\mbox{\boldmath${#1}$}}
\newcommand{\sect}[1]{\setcounter{equation}{0}\section{#1}}
\renewcommand{\theequation}{\arabic{section}.\arabic{equation}}
\newcommand{\app}{\setcounter{section}{0}
\setcounter{equation}{0} \renewcommand{\thesection}{APPENDIX
\Alph{section}}\renewcommand{\theequation}{\Alph{section}.
\arabic{equation}}}
\usepackage{graphicx}
\begin{document}

\title{Thermodynamic equilibrium and its stability for
microcanonical systems described by the Sharma-Taneja-Mittal
entropy}
\author{A.M. Scarfone$^{1,\,2}$}
\email{antonio.scarfone@polito.it}
\author{T. Wada$^{1,\,3}$}
\email{wada@ee.ibaraki.ac.jp} \affiliation{$^1$Dipartimento di
Fisica and $^2$Istituto Nazionale di Fisica della Materia (INFM),
Politecnico di Torino, Corso Duca degli Abruzzi 24, 10129 Torino,
Italy.\\$^3$Department of Electrical and Electronic Engineering,
Ibaraki University, Hitachi, Ibaraki, 316-8511, Japan}

\date{\today}
\begin{abstract}
It is generally assumed that the thermodynamic stability of
equilibrium state is reflected by the concavity of entropy.
We inquire, in the microcanonical picture, on the validity
of this statement for systems described by the
two-parametric entropy $S_{_{\kappa,\,r}}$ of
 Sharma-Taneja-Mittal. We analyze the
``composability'' rule for two statistically independent
systems, A and B, described by the entropy
$S_{_{\kappa,\,r}}$ with the same set of the deformation
parameters. It is shown that, in spite of the concavity of
the entropy, the ``composability'' rule modifies the
thermodynamic stability conditions of the equilibrium
state. Depending on the values assumed by the deformation
parameters, when the relation $S_{_{\kappa,\,r}}({\rm
A}\cup{\rm B})> S_{_{\kappa,\,r}}({\rm
A})+S_{_{\kappa,\,r}}({\rm B})$ holds (super-additive
systems), the concavity conditions does imply the
thermodynamics stability. Otherwise, when the relation
$S_{_{\kappa,\,r}}({\rm A}\cup{\rm
B})<S_{_{\kappa,\,r}}({\rm A})+S_{_{\kappa,\,r}}({\rm B})$
holds (sub-additive systems), the concavity conditions does
not imply the thermodynamical stability of the equilibrium
state.
\end{abstract}
 \maketitle

\sect{Introduction}

The MaxEnt principle of the thermodynamics, pioneered by Gibbs and
Janes \cite{Jaynes}, implies that, at equilibrium, both $d\,S=0$
and $d^2\,S<0$. The first of these conditions states that entropy
is
an extremum, whereas the second condition states that this extremum is a maximum.\\
It is well-known that from the second condition follow the
concavity conditions for the Boltzmann-Gibbs entropy which
are equivalent to the thermodynamic stability conditions
of its equilibrium distribution \cite{Callen,Chandler}.\\
Some interesting physical implications arise from the
thermodynamic stability conditions. For instance, the
positivity of the heat capacity assures that, for two
bodies in thermal contact and with different temperatures,
heat flows from the hot body to the cold one.

In the present days it is widely accepted that the
Boltzmann-Gibbs distribution represents only a special case
among the great diversity of statistical distributions
observed in nature. In many cases such distributions show
asymptotic long tails with a power-law behavior. Examples
include anomalous diffusion and Levy-flight
\cite{Zanette,Compte}, turbulence \cite{Beck},
self-gravitating systems \cite{Makino}, high $T_{_c}$
superconductivity \cite{Uys}, Bose-Einstein condensation
\cite{Tanatar}, kinetics of charge particles
\cite{Rossani},
biological systems \cite{Upadhyaya} and others.\\
To deal with such {\em anomalous} statistical systems, some
generalizations of the well-known Boltzmann-Gibbs entropy
have been advanced, with the purpose, by a way, to
incorporate the newly observed phenomenologies and, on the
other hand, to mimics the beautiful mathematical structure
of the standard thermostatistics theory
\cite{Abe,Kaniadakis0,Kaniadakis00}. A possible way to do
this is to replace the standard logarithm in the
Boltzmann-Gibbs entropy, $S(p)=-k_{_{\rm B}}\sum_i
p_{_i}\,\ln(p_{_i})$, with its generalized version
\cite{Naudts1,Naudts2}.

In this work we investigate the relationship between the
concavity conditions and the thermodynamic stability
conditions of the equilibrium distribution of a
conservative system,
with fixed energy and volume, described by a generalized entropy.\\
A preliminary investigation along this question can be
found in Refs. \cite{Ramshaw,Wada1,Wada2,Wada3}.\\ As a
working tool we employ the two-parameter entropy of
Sharma-Taneja-Mittal \cite{Taneja1,Taneja2,Mittal,Borges}.
Although, an entropy containing two free parameters could
sound unlike on the physical ground, the
Sharma-Taneja-Mittal entropy includes, as special cases,
some one-parameter entropies already proposed in
literature, like the Tsallis entropy
\cite{Tsallis,Tsallisweb}, the Abe entropy \cite{Abe1,Abe7}
and the Kaniadakis entropy \cite{Kaniadakis3,Kaniadakis2}.
Consequently, Sharma-Taneja-Mittal entropy enables us to
consider all this one-parameter entropies in a unified
scheme.

In Refs. \cite{Scarfone1,Scarfone2} it has been addressed
the question on the existence of a generalized trace-form
entropy
\begin{equation}
S(p)=-\sum_{i=1}^Wp_{_i}\Lambda(p_{_i}) \ ,\label{ent}
\end{equation}
(throughout this paper we use units with the Boltzmann
constant $k_{_{\rm B}}=1$) preserving unaltered the
epistemological structure of the standard statistical
mechanics. In Eq. (\ref{ent}) $\Lambda(x)$ is a deformed
logarithm \cite{Naudts2} replacing the standard one,
$\{p_{_i}\}_{_{i=1,\cdots,\,W}}$ is a discrete probability
distribution function and $W$ the number of microscopically
accessible states. By requiring that the entropy
(\ref{ent}) preserves the mathematical properties
physically motivated \cite{Kaniadakis2}, the following
differential-functional equation has been obtained
\begin{equation}
\frac{d}{d\,p_{_j}}\left[p_{_j}\,\Lambda(p_{_j})\right]
=\lambda\,\Lambda\left(\frac{p_{_j}}{\alpha}\right) \
.\label{condint}
\end{equation}
A physically suitable solution
$\Lambda(x)\equiv\ln_{(\kappa,\,r)}(x)$ can be written as
\begin{equation}
\ln_{_{\{\kappa,\,r\}}}(x)=x^r\,\frac{x^\kappa-
x^{-\kappa}}{2\,\kappa} \ ,\label{log1}
\end{equation}
which satisfies Eq. (\ref{condint}) with
\begin{equation}
\alpha=\left(\frac{1+r-\kappa}{1+r+\kappa}\right)^{1/2\,\kappa}
\ ,\label{alpha}
\end{equation}
and
\begin{equation}
\lambda=
\frac{\left(1+r-\kappa\right)}{\mbox{\raisebox{-1mm}
{$\left(1+r+\kappa\right)$}}}
^{\scriptscriptstyle{(r+\kappa)/2\,\kappa}}_
{\mbox{\raisebox{3.5mm}
{$\scriptscriptstyle{(r-\kappa)/2\,\kappa}$}}}
 \ .\label{lambda}
\end{equation}
Taking into account of Eq. (\ref{log1}), the generalized
entropy (\ref{ent}) assumes the form
\begin{eqnarray}
\nonumber
S_{_{\kappa,r}}(p)&=&-\sum_{i=1}^W(p_i)^{r+1}\,\frac{(p_{_i})^\kappa-
(p_{_i})^{-\kappa}}{2\,\kappa}\\
&=&-\sum_{i=1}^Wp_{_i}\,\ln_{_{\{\kappa,\,r\}}}(p_{_i})
 \ ,\label{entropyst}
\end{eqnarray}
which has been introduced for the first time in Refs.
\cite{Taneja1,Taneja2,Mittal} and
successively reconsidered in Ref. \cite{Borges}. \\
Eq. (\ref{entropyst}) mimics the expression of the
Boltzmann-Gibbs entropy by replacing the standard logarithm
$\ln(x)$ with the two-parametric deformed logarithm
$\ln_{_{\{\kappa,\,r\}}}(x)$.\\  The distribution obtained
by optimizing Eq. (\ref{entropyst}), under standard linear
energy constraint $\sum_ip_{_i}\,E_{_i}=\langle E\rangle$
and the normalization constraint $\sum_ip_{_i}=1$, assumes
the form
\begin{equation}
p_{_i}=\alpha\,\exp_{_{\{\kappa,\,r\}}}\left(-\frac{\beta}{\lambda}\,(E_{_i}-\mu)\right)
\ ,\label{dd}
\end{equation}
where $\exp_{_{\{\kappa,\,r\}}}(x)$ is the inverse function
of
$\ln_{_{\{\kappa,\,r\}}}(x)$, namely the deformed exponential.\\
Remarkably, Eq. (\ref{dd}) exhibits an asymptotic power law
behavior, with $p_{_i}\sim E_{_i}^{1/(r-\kappa)}$, for
large $E_{_i}$. This entropy possesses positivity,
continuity, symmetry, expandibility, decisivity,
maximality, concavity, and is Lesche stable whenever
$(\kappa,\,r)\in{\cal R}$, where the two dimensional region
$\cal R$ is defined by $-|\kappa|\leq r\leq|\kappa|$ for
$|\kappa|<1/2$ and $|\kappa|-1<r<1-|\kappa|$ for
$1/2\leq|\kappa|<1$.

We remark that the deformed logarithm (\ref{log1}) reduces
to the standard logarithm in the
$(\kappa,\,r)\rightarrow(0,\,0)$ limit
$(\ln_{_{\{0,\,0\}}}(x)\equiv\ln x)$ and, in the same
limit, Eq. (\ref{entropyst}) reduces to the Boltzmann-Gibbs
entropy. [Refer to the Appendix A for the main mathematical
properties of the deformed logarithm (\ref{log1})].

The plan of the paper is the following. In the next Section
2, we consider the Sharma-Taneja-Mittal entropy and its
distribution in the microcanonical framework. In Section 3
we derive the ``composability'' rule for two statistically
independent systems A and B with the same set of the
deformation parameters. In Section 4 we inquire on the
functional relationship between the Sharma-Taneja-Mittal
entropy and the definitions of temperature and pressure
obtained as equivalence relations at the equilibrium
configuration. In Section 5 we examine the thermodynamic
response produced by perturbing the system away from the
equilibrium. The perturbations are generated by
repartitioning the energy (heat transfer) and the volume
(work transfer) between the two systems A and B. According
to the MaxEnt principle such processes lead to a lower
entropy, provided that the whole system A$\cup$B was
initially in a stable equilibrium. By analyzing the signs
of the entropy changes for these processes we obtain the
corresponding thermodynamic stability conditions. Finally,
in section 6 we relate this results to some known
one-parameter cases. Concluding remarks are reported in
section 7. In Appendix A we give some mathematical
properties of the deformed logarithm, and Appendix B deals
with the sketch of some proofs.

\sect{Microcanonical Sharma-Taneja-Mittal entropy}

According to the MaxEnt principle, the equilibrium distribution is
the one that maximizes the entropy under the constraints imposed
on the probability
distribution.\\
In the microcanonical picture, the system has fixed total energy
$E$ and volume $V$, and the distribution
$p\equiv\{p_{_i}\}_{_{i=1,\,\cdots,\,W}}$ is obtained by
optimizing the entropy (\ref{entropyst}) under the only constraint
on the normalization
\begin{equation}
\sum_{i=1}^W p_{_i}=1 \ .\label{norm}
\end{equation}

Thus, we have to deal with the variational problem
\begin{equation}
\frac{\delta}{\delta\,p_{_j}}\,\left(S_{_{\kappa,\,r}}(p)
-\gamma\sum_{i=1}^Wp_{_i} \right)=0 \ ,\label{var}
\end{equation}
where $\gamma$ is the Lagrange multiplier associated with
the constraint (\ref{norm}). By taking into account Eqs.
(\ref{ent})-(\ref{condint}), it follows
\begin{equation}
\lambda\,\ln_{_{\{\kappa,\,r\}}}\left(\frac{p_{_j}}{\alpha}\right)
+\gamma=0 \ ,\label{eq1}
\end{equation}
and by means of the deformed exponential
$\exp_{_{\{\kappa,\,r\}}}(x)$, we obtain
\begin{equation}
p_{_j}=\alpha\,\exp_{_{\{\kappa,\,r\}}}
\left(-\frac{\gamma}{\lambda}\right) \ .\label{dist}
\end{equation}
Since this distribution does not depend on the index $j$,
according to Eq. (\ref{norm}), it takes the form
\begin{equation}
p_{_j}={1\over W(E,\,V)} \ ,\hspace{10mm}{\rm
with}\hspace{10mm}j=1,\,\cdots,\,W \ ,\label{distr}
\end{equation}
where we took into account that the number of accessible states
$W(E,\,V)$ is a
function of the energy $E$ and the volume $V$ of the system.\\
By substituting Eq. (\ref{distr}) into Eq.
(\ref{entropyst}) we obtain its expression in the
microcanonical picture
\begin{eqnarray}
\nonumber
S_{_{\kappa,r}}(E,\,V)&=&-\ln_{_{\{\kappa,\,r\}}}\Bigg({1\over
W(E,\,V)}\Bigg)\\&=&\ln_{_{\{\kappa,\,-r\}}}\Big(W(E,\,V)\Big)
\ . \label{Boltzmann}
\end{eqnarray}
This is evocative of the well-known Boltzmann formula
$S=\ln(W)$, which is indeed recovered in the
$(\kappa,\,r)\to(0,\,0)$ limit.

We observe that the concavity of the function
$\ln_{_{\{\kappa,\,r\}}}(x)$ w.r.t. its argument $x$ does not
necessarily imply the concavity of the entropy (\ref{Boltzmann})
w.r.t $E$ and $V$.\\
The concavity conditions for the given problem follow from
the analysis of the sign of the eigenvalues of the Hessian
matrix associated to Eq. (\ref{Boltzmann}). In particular,
by requiring that the following quadratic form is negative
definite \cite{Hancock}
\begin{equation}
\phi({\bfm y};\,E,\,V)=\frac{\partial^2S_{_{\kappa,r}}}{\partial
E^2}\,y_{_{\rm E}}^2 +2\,\frac{\partial^2S_{_{\kappa,r}}}{\partial
E\,\partial V}\,y_{_{\rm E}} \,y_{_{\rm
V}}+\frac{\partial^2S_{_{\kappa,r}}}{\partial V^2}\,y_{_{\rm V}}^2
\ ,\label{quad}
\end{equation}
for any arbitrary vector ${\bfm y}\equiv(y_{_{\rm E}},\,y_{_{\rm
V}})$, we obtain the following relations
\begin{equation}
\frac{\partial^2S_{_{\kappa,r}}}{\partial E^2}<0 \ ,\label{st1}
\end{equation}
and
\begin{equation}
\frac{\partial^2S_{_{\kappa,r}}}{\partial
E^2}\,\frac{\partial^2S_{_{\kappa,r}}}{\partial V^2}-
\left(\frac{\partial^2S_{_{\kappa,r}}}{\partial E\,\partial
V}\right)^2>0 \ ,\label{st2}
\end{equation}
stating the concavity conditions for the entropy
(\ref{Boltzmann}).


\sect{Composed systems}

Let us consider two systems A and B described by the entropy
(\ref{Boltzmann}),
with the same set of the deformation parameters.\\
We denote with $W_{\rm A}\equiv W(E_{\rm A},\,V_{\rm A})$
and $W_{\rm B}\equiv W(E_{\rm B},\,V_{\rm B})$ the number
of accessible states of the two systems A and B
respectively and hypothesize a statistical independence of
A and B, in the sense that the number of accessible sates
$W_{{\rm A}\cup{\rm B}}\equiv W(E_{{\rm A}\cup {\rm
B}},\,V_{{\rm A}\cup{\rm B}})$ of the composed system
A$\cup$B is given by $W_{{\rm A}\cup{\rm B}}={W_{\rm A}}\,{W_{\rm B}}$.\\
In Ref. \cite{Abe2} the most general form of pseudoadditivity of
composable entropies, as prescribed by the existence of
equilibrium, has been obtained. The main result reads
\begin{equation}
H[S({\rm A}\cup{\rm B})]=H[S({\rm A})]+H[S({\rm
B})]+\lambda\,H[S({\rm A})]\,H[S({\rm B})] \
,\label{composability}
\end{equation}
where $H(x)$ is a certain differentiable function,
$\lambda$ denotes the set of deformation parameters, whilst
$S({\rm A})$, $S({\rm B})$ and $S({\rm{A}\cup{\rm B}})$ are
the entropies of the
systems A, B and A$\cup$B, respectively.\\
It is easy to show that Eq. (\ref{composability}) is fulfilled by
the entropy (\ref{Boltzmann}) if we define
\begin{eqnarray}
\nonumber
H[S_{_{\kappa,\,r}}(W)]&=&-S_{_{\kappa,\,r}}(W)\left[\exp_{_{\{\kappa,\,r\}}}
\left(-S_{_{\kappa,\,r}}(W)\right)\right]^{-r-\kappa}\\
&=&W^{r+\kappa}\,\ln_{_{\{\kappa,\,r\}}}\left({1\over W}\right) \
.\label{h}
\end{eqnarray}
In fact, by using Eq. (\ref{rl2}) given in Appendix A, we have
\begin{eqnarray}
\nonumber & &\left(W_{\rm A}\,W_{\rm
B}\right)^{r+\kappa}\,\ln_{_{\{\kappa,\,r\}}}\left({1\over
W_{\rm A}\,W_{\rm B}}\right)\\
\nonumber&=&(W_{\rm
A})^{r+\kappa}\,\ln_{_{\{\kappa,\,r\}}}\left({1\over W_{\rm
A}}\right)+(W_{\rm
B})^{r+\kappa}\,\ln_{_{\{\kappa,\,r\}}}\left({1\over W_{\rm
B}}\right)\\ \nonumber& &-2\,\kappa\,\left[(W_{\rm
A})^{r+\kappa}\,\ln_{_{\{\kappa,\,r\}}}\left({1\over W_{\rm
A}}\right)\right]\\
&\times&\left[(W_{\rm
B})^{r+\kappa}\,\ln_{_{\{\kappa,\,r\}}}\left({1\over W_{\rm
B}}\right)\right] \ , \label{comp1}
\end{eqnarray}
which has the same structure of Eq. (\ref{composability}) with
$H[S_{_{\kappa,\,r}}]$ given in Eq. (\ref{h}) and $\lambda=-2\,\kappa$.\\
After multiplying Eq. (\ref{comp1}) by $\left(W_{\rm A}\, W_{\rm
B}\right)^{-r-\kappa}$ and recalling the invariance of the entropy
(\ref{Boltzmann}) under the interchange of
$\kappa\leftrightarrow-\kappa$, Eq. (\ref{comp1}) becomes
\begin{equation}
S_{_{\kappa,\,r}}({\rm A}\cup{\rm B})=S_{_{\kappa,\,r}}({\rm
A})\,{\mathcal I}_{_{\kappa,\,r}}({\rm B})+{\mathcal
I}_{_{\kappa,\,r}}({\rm A})\,S_{_{\kappa,\,r}}({\rm B}) \ ,
\label{comp3}
\end{equation}
where the function ${\mathcal I}_{\kappa,\,r}(p)$, for a given
distribution function $p=\{p_{_i}\}_{_{i=1,\,\cdots,\,W}}$, is
defined by
\begin{equation}
{\mathcal I}_{_{\kappa,\,r}}(p)=\sum_{i=1}^W(p_{_i})^{r+1}
\frac{(p_{_i})^\kappa+(p_{_i})^{-\kappa}}{2} \ ,
\end{equation}
with ${\mathcal I}_{_{0,\,0}}(p)=1$ and reduces, for the uniform
distribution (\ref{distr}), to
\begin{equation}
{\mathcal I}_{_{\kappa,\,r}}\left({1\over W(E,\,V)}\right)=\frac{
\left[W(E,\,V)\right]^{-r-\kappa}+\left[W(E,\,V)\right]^{-r+\kappa}}{2}
\ .\label{i}
\end{equation}
We remark that Eq. (\ref{i}) actually is a function of the entropy
(\ref{Boltzmann}) through the relation
\begin{equation}
{\mathcal I}_{_{\kappa,\,r}}(x)=\kappa
\,S_{_{\kappa,\,r}}(x)+\left[\exp_{_{\{\kappa,\,r\}}}
\left(-S_{_{\kappa,\,r}}(x)\right)\right]^{r+\kappa} \
.\label{qmic}
\end{equation}

It is worthy to observe that Eq. (\ref{comp3}) still holds, for a
canonical distribution, also if the entropy (\ref{entropyst}), in
this case, does not satisfy the criteria dictated by Eq.
(\ref{composability}).

Eq. (\ref{comp3}) expresses the ``composability''
properties for a system described by the entropy
(\ref{Boltzmann}), and, in the $(\kappa,\,r)\to(0,\,0)$
limit, we recover the well-known additivity rule of the
Boltzmann entropy $S({\rm A}\cup{\rm B})=S({\rm A})+S({\rm
B})$.\\
In the following we analyze in more details this equation.\\
According to the results given in the Appendix A, when
$(\kappa,\,r)\in{\cal R}\big|_{r\leq0}$ it follows ${\mathcal
I}_{_{\kappa,\,r}}\left(1/W\right)>1$ for $W>1$. Consequently,
from Eq. (\ref{comp3}) we obtain
\begin{equation}
S_{_{\kappa,\,r}}({\rm A}\cup{\rm B})>S_{_{\kappa,\,r}}({\rm
A})+S_{_{\kappa,\,r}}({\rm B}) \ ,\label{super}
\end{equation}
and the entropy (\ref{Boltzmann}) exhibits a super-additive behavior.\\
The analysis of the Eq. (\ref{comp3}) becomes more complicated in
the complementary region ${\cal R}\big|_{r>0}$. In fact, for $r >
0$ there exists a threshold point $W_{\rm t}(\kappa,\,r)>1$, which
is defined by
\begin{equation}
{\mathcal I}_{_{\kappa,\,r}}\left({1\over W_{\rm t}}\right)=1 \
,\label{tnew}
\end{equation}
so that ${\mathcal I}_{_{\kappa,\,r}}\left(1/W\right)\leq1$ when
$1<W\leq W_{\rm t}$ whereas ${\mathcal
I}_{_{\kappa,\,r}}\left(1/W\right)>1$ when $W>W_{\rm t}$.
Consequently, for $(\kappa,\,r)\in{\cal R}\big|_{r>0}$ we have a
sub-additive behavior
\begin{equation}
S_{_{\kappa,\,r}}({\rm A}\cup{\rm B})< S_{_{\kappa,\,r}}({\rm
A})+S_{_{\kappa,\,r}}({\rm B}) \ ,
\end{equation}
when both $1<W_{\rm A}<W_{\rm t}$ and $1<W_{\rm B}<W_{\rm t}$,
whereas the super-additive behavior (\ref{super}) is recovered
whenever $W_{\rm A}>W_{\rm t}$ and $W_{\rm B}>W_{\rm t}$. In the
intermediate situation $1<W_{\rm A}<W_{\rm t}$ and $W_{\rm
B}>W_{\rm t}$, or $W_{\rm A}>W_{\rm t}$ and $1<W_{\rm B}<W_{\rm
t}$, the character of the composition law is not well determined,
depending on the values of $W_{\rm A}$ and
$W_{\rm B}$.\\
Thus, for $(\kappa,\,r)\in{\cal R}\big|_{r>0}$ the value of the
entropy of a composed system A$\cup$B, w.r.t. the sum of the
entropies of the two separate systems A and B, depends on the {\em
size} of the two systems \footnote{Here {\em size} is used to
indicate the value $W$ of the accessible states of the system.}.
{\em Small} systems exhibit a sub-additive behavior, which becomes
super-additive when both the systems grow over the threshold point $W_{\rm t}$.\\
As consequence, super-additivity behavior emerges with {\em
larger} systems.\\ We observe that the threshold point $W_{\rm t}$
becomes larger and larger, for $r\to\kappa$, according to
\begin{equation} \lim_{r\to\kappa}W_{\rm
t}(\kappa,\,r)\to\infty \ .
\end{equation}
As a consequence, ${\mathcal I}_{_{\kappa,\,\kappa}}\leq1$ and the
entropy $S_{_{\kappa,\,\kappa}}({\rm A}\cup{\rm B})$ has always a
sub-additive behavior.

\sect{Thermal and mechanical equilibrium}

Possible definitions of temperature and pressure, in the
construction of a generalized framework of thermodynamics, have
been proposed in Refs. \cite{Abe3,Abe5,Abe6} through the study of
the equilibrium configuration.\\
Such method can be successfully applied to the generalized entropy
under inspection.\\ We assume that both energy and volume are
additive quantities, i.e., $E_{\rm A\cup B}=E_{\rm A}+E_{\rm B}$
and $V_{\rm A\cup B}=V_{\rm A}+V_{\rm B}$. A different approach,
by utilizing nonadditive energy and volume, within the framework
of
nonextensive statistical mechanics, has been explored in Ref. \cite{Wang}.\\
Let us consider an isolated system A$\cup$B composed by two,
statistically independent, systems A and B in contact through an
ideal wall. The wall permits transfer of energy (heat) and/or
volume (work) between the two systems but is adiabatic with
respect to any other interaction.\\
We suppose that the system, initially at the thermal and
mechanical equilibrium, undergoes a small fluctuation of energy
and volume between A and B. According to the MaxEnt principle the
variation of the entropy evaluated at the first order in $\delta
E$ and $\delta V$ must vanish
\begin{eqnarray}
&&\delta S_{\kappa,\,r}({\rm A}\cup{\rm B})=0 \ ,\label{ent1}
\end{eqnarray}
where
\begin{eqnarray}
&&\delta (E_{\rm A}+E_{\rm B})=0 \ ,\label{en}\\
&&\delta (V_{\rm A}+V_{\rm B})=0 \ .\label{vo}
\end{eqnarray}
From Eq. (\ref{comp3}) we obtain (see Appendix B)
\begin{equation}
{1\over{\mathcal I}_{_{\kappa,\,r}}-r\,S_{_{\kappa,\,r}}}
\,{\partial\,S_{_{\kappa,\,r}}\over\partial\,E}\Bigg|_{\rm
A}={1\over{\mathcal I}_{_{\kappa,\,r}}-r\,S_{_{\kappa,\,r}}}\,
{\partial\,S_{_{\kappa,\,r}}\over\partial\,E}\Bigg|_{\rm B} \
,\label{0law}
\end{equation}
and
\begin{equation}
{1\over{\mathcal
I}_{_{\kappa,\,r}}-r\,S_{_{\kappa,\,r}}}\,{\partial\,S_{_{\kappa,\,r}}
\over\partial\,V}\Bigg|_{\rm A}={1\over{\mathcal
I}_{_{\kappa,\,r}}-r\,S_{_{\kappa,\,r}}}\,
{\partial\,S_{_{\kappa,\,r}}\over\partial\,V}\Bigg|_{\rm B} \
.\label{mech}
\end{equation}
Actually Eqs. (\ref{0law}) and (\ref{mech}) state the
analytical formulation of the zeroth law of the
thermodynamics for the system under inspection and define,
as equivalence relations, modulo of a multiplicative
constant, the temperature
\begin{equation}
{1\over T}={1\over{\mathcal
I}_{_{\kappa,\,r}}-r\,S_{_{\kappa,\,r}}}\,{\partial\,S_{_{\kappa,\,r}}
\over\partial\,E} \ ,\label{temp}
\end{equation}
and the pressure
\begin{equation}
P={T\over{\mathcal
I}_{_{\kappa,\,r}}-r\,S_{_{\kappa,\,r}}}\,{\partial\,S_{_{\kappa,\,r}}
\over\partial\,V} \ ,\label{pres}
\end{equation}
which, accounting Eq. (\ref{qmic}), are given only through the
entropy $S_{_{\kappa,\,r}}$.\\ The standard relations of the
classical thermodynamics
\begin{equation}
{1\over T}=\frac{\partial\,S}{\partial\,E} \ ,\label{tc}
\end{equation}
and
\begin{equation}
P=T\,\frac{\partial\,S}{\partial\,V} \ ,\label{pc}
\end{equation}
are recovered in the $(\kappa,\,r)\to (0,\,0)$ limit.

It is worth to observe that, by using Eq. (\ref{Boltzmann}), Eqs.
(\ref{temp})-(\ref{pres}) can be written as
\begin{equation}
{1\over T}=\frac{\partial}{\partial\,E}\,\ln(W) \ ,\label{t}
\end{equation}
and
\begin{equation}
P=T\,\frac{\partial}{\partial\,V}\,\ln(W) \ ,\label{p}
\end{equation}
which are positive definite quantities if $W(E,\,V)$ is a
monotonic increasing function with respect to both $E$ and $V$.

In Refs. \cite{Toral,Velazquez} has been noted that in the
microcanonical framework of the Tsallis' thermostatistics , the
definitions of $T$ and $P$, obtained through the study of the
equilibrium configuration, lead to expressions which coincide with
those obtained by using the standard Boltzmann formalism of
statistical mechanics. This results still hold in presence of the
entropy (\ref{Boltzmann}), as can be seen from Eqs. (\ref{t}) and
(\ref{p}), which define temperature and pressure in function of
$W$ and coincide with the standard definitions adopted in the
Boltzmann theory.

\sect{Thermodynamic stability}

In this section we examine the thermodynamic response produced by
perturbing the system which is assumed initially in equilibrium.
By analyzing the signs of thermodynamic changes, we obtain the
thermodynamic stability conditions. \\
Let us consider a small perturbation of the system through a
transfer of an amount of energy and/or volume between A and B:
$S_{_{\kappa,\,r}}({\rm A}\cup{\rm B})\rightarrow
S_{_{\kappa,\,r}}(({\rm A}+\delta{\rm A})\cup({\rm B}+\delta{\rm
B}))$.\\ According to the MaxEnt principle, such a perturbation
leads the system in a new state with a lower entropy
\begin{equation}
S_{_{\kappa,\,r}}({\rm A}\cup{\rm B})> S_{_{\kappa,\,r}}(({\rm
A}+\delta{\rm A})\cup({\rm B}+\delta{\rm B})) \ .\label{fluc}
\end{equation}
In Eq. (\ref{fluc}) we denote $S_{_{\kappa,\,r}}({\rm A}\cup{\rm
B})\equiv S_{_{\kappa,\,r}}((E_{\rm A}+E_{\rm B},\,V_{\rm
A}+V_{\rm B}))$ and $S_{_{\kappa,\,r}}(({\rm A}+\delta{\rm
A})\cup({\rm B}+\delta{\rm B}))\equiv S_{_{\kappa,\,r}}((E_{\rm
A}+\delta E_{\rm A},\,V_{\rm A}+\delta V_{\rm A})\cup(E_{\rm
B}+\delta E_{\rm B},\,V_{\rm B}+\delta V_{\rm B}))$ where $\delta
E_{\rm A}=-\delta E_{\rm B}\equiv \delta E$ and $\delta V_{\rm
A}=-\delta V_{\rm B}\equiv \delta V$.\\ Recalling Eq.
(\ref{comp3}), Eq. (\ref{fluc}) can be written in
\begin{eqnarray}
\nonumber &&S_{_{\kappa,\,r}}({\rm A})\,{\mathcal
I}_{_{\kappa,\,r}}({\rm B})+{\mathcal
I}_{_{\kappa,\,r}}({\rm A})\,S_{_{\kappa,\,r}}({\rm B})>\\
\nonumber & &S_{_{\kappa,\,r}}\left({\rm A}+\delta{\rm
A}\right)\,{\mathcal I}_{_{\kappa,\,r}}({\rm B}+\delta {\rm
B})\\&+&{\mathcal I}_{_{\kappa,\,r}}\left({\rm
A}+\delta{\rm A}\right) \,S_{_{\kappa,\,r}}\left({\rm
B}+\delta{\rm B}\right) \ ,\label{fluct}
\end{eqnarray}
and after expanding the r.h.s of Eq. (\ref{fluct}) up to the
second order in $\delta E$ and $\delta V$, we obtain (see the
Appendix B)
\begin{eqnarray}
\nonumber& &{1\over2}\left({\mathcal I}_{_{\kappa,\,r}}-
r\,S_{_{\kappa,\,r}}\right)\Bigg|_{{\rm A}\cup{\rm
B}}\\
\nonumber& &\Bigg[{{\mathcal S}_{_{\rm EE}}\,(\delta
E)^2+2\,{\mathcal S}_{_{\rm EV}}\,\delta E\,\delta
V+{\mathcal S}_{_{\rm VV}}\,(\delta V)^2\over{\mathcal
I}_{_{\kappa,\,r}}-r\,S_{_{\kappa,\,r}}}\Bigg|_{\rm A}\\
\nonumber+& &{{\mathcal S}_{_{\rm EE}}\,(\delta
E)^2+2\,{\mathcal S}_{_{\rm EV}}\,\delta E\,\delta
V+{\mathcal S}_{_{\rm VV}}\,(\delta V)^2\over{\mathcal
I}_{_{\kappa,\,r}}-r\,S_{_{\kappa,\,r}}}\Bigg|_{\rm
B}\Bigg]<0 \ ,\\ \label{fluct1}
\end{eqnarray}
where we have posed
\begin{equation}
{\mathcal S}_{_{\rm XY}}=\frac{\,\,\partial^2
S_{_{\kappa,\,r}}}{\partial X\,\partial
Y}-\frac{(\kappa^2+r^2)\,S_{_{\kappa,\,r}}-2\,r\,{\mathcal
I}_{_{\kappa,\,r}}}{({\mathcal
I}_{_{\kappa,\,r}}-r\,S_{_{\kappa,\,r}})^2}\,\frac{\,\partial
S_{_{\kappa,\,r}}}{\partial X}\,\frac{\,\partial
S_{_{\kappa,\,r}}}{\partial Y} \ .
\end{equation}
Eq. (\ref{fluct1}) is fulfilled if the following inequalities
\begin{eqnarray}
&&{\mathcal S}_{_{\rm EE}}<0 \ ,\label{cs1}\\
&& {\mathcal S}_{_{\rm EE}}\,{\mathcal S}_{_{\rm VV}}-{\mathcal
S}_{_{\rm EV}}^2>0 \ ,\label{cs2}
\end{eqnarray}
are separately satisfied by both systems A and B.\\
Remark that Eqs. (\ref{cs1})-(\ref{cs2}) have the same
structure of Eq. (\ref{st1})-(\ref{st2}).\\ Explicitly,
Eqs. (\ref{cs1})-(\ref{cs2}) read
\begin{eqnarray}
\frac{\partial^2S_{\kappa,\,r}}{\partial E^2}<{\mathcal
A}_{_{\kappa,\,r}}\,
\left(\frac{\partial\,S_{\kappa,\,r}}{\partial\,E}\right)^2 \ ,
\label{tsc2}
\end{eqnarray}
and
\begin{eqnarray}
&&\frac{\,\,\partial^2 S_{_{\kappa,\,r}}}{\partial^2
E}\,\frac{\,\,\partial^2 S_{_{\kappa,\,r}}}{\partial^2
V}-\left(\frac{\,\,\partial^2 S_{_{\kappa,\,r}}}{\partial
E\,\partial V}\right)^2>{\mathcal A}_{_{\kappa,\,r}}\,{\mathcal
B}_{_{\kappa,\,r}} \ ,\label{tsc3}
\end{eqnarray}
where
\begin{eqnarray}
{\mathcal
A}_{_{\kappa,\,r}}=\frac{(\kappa^2+r^2)\,S_{\kappa,\,r}-2\,r\,{\mathcal
I}_{_{\kappa,\,r}}}{\left({\mathcal
I}_{_{\kappa,\,r}}-r\,S_{\kappa,\,r}\right)^2} \ ,\label{a}
\end{eqnarray}
and
\begin{eqnarray}
\nonumber {\mathcal
B}_{_{\kappa,\,r}}\!\!&=&\!\!\left(\frac{\partial^2
S_{_{\kappa,\,r}}}{\partial
E^2}\right)^{\!\!-1}\!\!\left\{\left(\frac{\partial^2
S_{_{\kappa,\,r}}}{\partial E^2}\frac{\partial
S_{_{\kappa,\,r}}}{\partial V}-\frac{\partial^2
S_{_{\kappa,\,r}}}{\partial E\,\partial V}\frac{\partial
S_{_{\kappa,\,r}}}{\partial
E}\right)^{\!\!2}\right.\\
\nonumber &+&\!\!\left.\left(\frac{\partial
S_{_{\kappa,\,r}}}{\partial
E}\right)^{\!\!2}\left[\frac{\partial^2S_{_{\kappa,\,r}}}{\partial
E^2}\,\frac{\partial^2S_{_{\kappa,\,r}}}{\partial V^2}-
\left(\frac{\partial^2 S_{_{\kappa,\,r}}}{\partial
E\,\partial V}\right)^2\right]\right\} \ .\\ \label{b}
\end{eqnarray}
In particular, the quantity ${\mathcal B}_{_{\kappa,\,r}}$ is
negative definite for a concave entropy, as a consequence of Eqs. (\ref{st1})-(\ref{st2}).\\
From Eqs. (\ref{cs1}) and (\ref{cs2}) it is trivial to
obtain the further inequality ${\mathcal S}_{\rm vv}<0$,
which can be written in
\begin{eqnarray}
\frac{\partial^2S_{\kappa,\,r}}{\partial V^2}<{\mathcal
A}_{_{\kappa,\,r}}\,
\left(\frac{\partial\,S_{\kappa,\,r}}{\partial\,V}\right)^2 \ .
\label{tsc4}
\end{eqnarray}
Eqs. (\ref{tsc2}), (\ref{tsc3}) and (\ref{tsc4}) are the
announced thermodynamic stability conditions for the family
of entropies given in Eq. (\ref{entropyst}), and reduce to
Eq. (\ref{st1})-(\ref{st2}) in the $(\kappa,\,r)\to(0,\,0)$
limit. Depending on the sign of the function ${\cal
A}_{_{\kappa,\,r}}$ this inequalities are satisfy if the
concavity conditions does.

We observe that the equation ${\cal A}_{_{\kappa,\,r}}({\widetilde
x}_{\rm t})=0$ has the unique solution given by
\begin{equation}
{\widetilde x}_{\rm
t}=\left(\frac{\kappa-r}{\kappa+r}\right)^{1/\kappa} \ .\label{wt}
\end{equation}
By inspection it follows that ${\cal A}_{_{\kappa,\,r}}(x)>0$ when
$0\leq x\leq \widetilde x_{\rm t}$ whilst ${\cal
A}_{_{\kappa,\,r}}(x)<0$ when $\widetilde x_{\rm t}\leq
x\leq+\infty$. In the parametric region  $(\kappa,\,r)\in{\cal
R}\big|_{r\leq0}$, accounting for Eq. (\ref{wt}), it follows that
$1/{\widetilde W}_{\rm t}={\widetilde x}_{\rm t}>1$. Thus, being
$W\geq1$, it follows $1/W<1<1/\widetilde W_{\rm t}$ so that ${\cal
A}_{_{\kappa,\,r}}>0$ and consequently
Eqs. (\ref{tsc2}), (\ref{tsc3}) and (\ref{tsc4})
are fulfilled if the concavity conditions are accomplished.\\
Differently, in the parametric region $(\kappa,\,r)\in{\cal
R}_{r>0}$, we have $1/{\widetilde W}_{\rm t}<1$. In this
case ${\cal A}_{_{\kappa,\,r}}\geq0$ when $W\geq\widetilde
W_{\rm t}$. As a consequence we obtain that the concavity
conditions imply the thermodynamic stability conditions if
and only if both $W_{\rm A}\geq\widetilde W_{\rm t}$ and
$W_{\rm B}\geq\widetilde
W_{\rm t}$ are satisfied.\\
At this point we observe that, when $r>0$
\begin{equation}
{\mathcal I}_{_{\kappa,\,r}}\left({1\over{\widetilde W}_{\rm
t}}\right)<1 \ ,
\end{equation}
so that
\begin{equation}
{\widetilde W}_{\rm t}<W_{\rm t} \ ,
\end{equation}
where $W_{\rm t}$ is the threshold point defined in Eq.
(\ref{tnew}), and the system becomes super-additive when
the number of accessible states $W_{\rm A}$ and $W_{\rm B}$
beyond $W_{\rm t}$. Thus, we can conclude that, whenever
the system exhibits a super-additive behavior, the
concavity conditions are sufficient to guarantee the
thermodynamic stability of the equilibrium configuration.

\sect{Examples}

In this section we specify our results to some
one-parameter entropies, already known in literature, and
belonging to the family of the Sharma-Taneja-Mittal
entropy.\\ In figure 1 we depict the log-linear plots for
the three one-parameter entropies discussed in this
section, for different values of the deformation parameter.
The solid line shows the Boltzmann entropy.\\

\begin{figure}[ht]
\includegraphics[width=.5\textwidth]{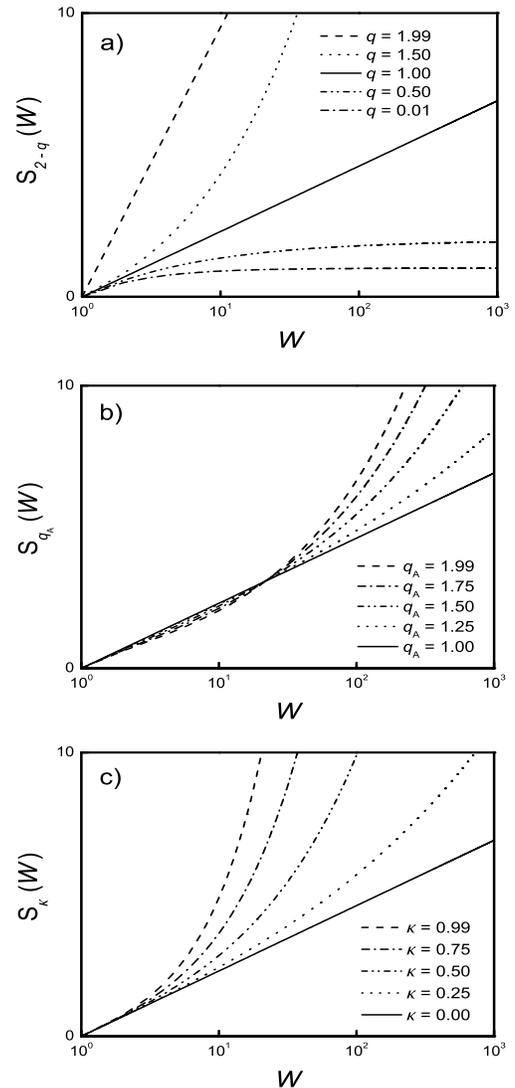}
\caption{Log-linear plot of some one-parameter entropies
(in arbitrary units) belonging to the family
(\ref{Boltzmann}) for several values of the deformation
parameter: $({\bfm A})$ Tsallis entropy, \cite{Tsallis};
$({\bfm B})$ Abe entropy, \cite{Abe1}; and $({\bfm C})$
Kaniadakis entropy, \cite{Kaniadakis2}. The solid curves
show the Boltzmann entropy. \label{fig:entropies} }
\end{figure}

\subsection{Tsallis entropy}

As a first example, we consider the Tsallis entropy
\cite{Tsallis}
\begin{equation}
S_{_{2-q}}=-\sum_{i=1}^W\frac{p^{2-q}_i-p_i}{q-1}=-\sum_{i=1}^Wp_{_i}\,\ln_{_q}\Big(p_{_i}\Big)
\ ,\label{tsa}
\end{equation}
with $0<q<2$, which follows from Eq. (\ref{entropyst}) by
posing $r=\pm|\kappa|$ and introducing the parameter
$q=1\mp2\,|\kappa|$. We remark that Eq. (\ref{tsa}) differs
from the usual definition adopted in the Tsallis framework
which is recovered by replacing $q\to2-q$.\\ In Eq.
(\ref{tsa}) the $q$-logarithm, $\ln_{_q}(x)$, is defined by
\begin{equation}
\ln_{_q}(x)=\frac{x^{1-q}-1}{1-q} \ ,\label{qlog}
\end{equation}
whereas, its inverse function, namely the $q$-exponential,
is given by
\begin{equation}
\exp_{_q}(x)=\left[1+(1-q)\,x\right]^{1/(1-q)} \
.\label{qexp}
\end{equation}
Both Eqs. (\ref{qlog}) and (\ref{qexp}) fulfill the
relations
\begin{eqnarray}
&&\exp_{_q}(x)\,\exp_{_q}(y)=\exp_{_q}(x\oplus_q y)
\ ,\label{qe}\\
&&\ln_{_q}(x\,y)=\ln_{_q}(x)\oplus_q\ln_{_q}(y) \
,\label{ql}
\end{eqnarray}
where the $q$-deformed sum, introduced in
\cite{Borges1,Wang1}, is defined as
\begin{equation}
x\oplus_qy=x+y+(1-q)\,x\,y \ .\label{qsum}
\end{equation}
Eqs. (\ref{qe}) and (\ref{ql}) reduce, in the $q\to1$
limit, to the well known standard relations
$\exp(x)\,\exp(y)=\exp(x+y)$ and $\ln(x\,y)=\ln(x)+\ln(y)$,
respectively,
according to $x\oplus_1y=x+y$.\\
After its introduction in 1988, Tsallis entropy has been
largely applied, as a paradigm, in the study of complex
systems having a probability distribution function with a
power law behavior in the tail. Typically, these systems
are characterized by long-range interactions or long time
memory effects that establish a space-time interconnection
by the parts which causes a strong interdependence and the
existence of a rich structure over several scales. All
these induce correlations between the parts of the system
which gives origin to a dynamical equilibrium rather than a
static equilibrium: the system remains in a metastable
configuration that could persist for a long period of time
as compared with the characteristic time scale of the
underlying microscopic dynamical process.\\ As it is known,
the Tsallis entropy exhibits many interesting proprieties
which make it a suitable substitute of the Boltzmann Gibbs
entropy in the study of these anomalous systems. Among them
we recall that it is concave for $q>0$, Lesche stable
\cite{Abe8}, a basic propriety which must be satisfied in
order to represent a well defined physical observable
\cite{Lesche}, and fulfill the Pesin identity \cite{Zheng}
stating a relation between the sensitivity to the initial
conditions and the (finite) entropy production per unit
time. Many other physical proprieties about the Tsallis
entropy can be found in \cite{Gelmann}.

In the microcanonical picture, with a uniform distribution
$p_{_i}=1/W$, Eq. (\ref{tsa}) reduces to
\begin{equation}
S_{_{2-q}}=-\ln_{_q}\Bigg({1\over W}\Bigg) \ ,\label{mts}
\end{equation}
and we introduce the function ${\mathcal I}_{_{2-q}}$
which, according to Eq. (\ref{qmic}), can be expressed as a
linear function of the entropy
\begin{equation}
{\mathcal I}_{_{2-q}}={1\over2}\,(q-1)\,S_{_{2-q}}+1 \
.\label{iq}
\end{equation}
From Eq. (\ref{iq}) it readily follows that ${\mathcal
I}_{_{2-q}}>1$ for $q>1$ and ${\mathcal I}_{_{2-q}}<1$ for
$q<1$, and from Eq. (\ref{comp3}) we obtain the well-known
``composability'' rule
\begin{eqnarray}
\nonumber & &S_{_{2-q}}\left({\rm A}\cup{\rm
B}\right)=S_{_{2-q}}\left({\rm
A}\right)+S_{_{2-q}}\left({\rm
B}\right)\\&+&(q-1)\,S_{_{2-q}}\left({\rm
A}\right)S_{_{2-q}}\left({\rm B}\right) \ ,\label{ta}
\end{eqnarray}
which shows that $S_{_{2-q}}$ is sub-additive for
$q\in(0,\,1)$ and super-additive for $q\in(1,\,2)$. Remark
that Eq. (\ref{ta}) can be readily obtained also from the
proprieties (\ref{ql}) of the $q$-logarithm.\\
Temperature and pressure are defined through \cite{Abe5,Abe6}
\begin{equation}
{1\over T}={1\over
1+(q-1)\,S_{_{2-q}}}\,\frac{\partial\,S_{_{2-q}}}{\partial\,E}
\ ,
\end{equation}
\begin{equation}
P={T\over1+(q-1)\,S_{_{2-q}}}\,\frac{\partial\,S_{_{2-q}}}{\partial\,V}
\ ,
\end{equation}
respectively, whereas the thermodynamic stability
conditions are obtained through Eqs.
(\ref{tsc2})-(\ref{tsc4}) and read \cite{Wada2,Wada3}
\begin{eqnarray} & &\frac{\partial^2S_{_{2-q}}}{\partial E^2}<{\cal
A}_{_{2-q}}\,\left(\frac{\partial S_{_{2-q}}} {\partial
E}\right)^2 \ ,\label{tscq1}\\ & &\nonumber\\&
&\frac{\partial^2 S_{_{2-q}}}{\partial
E^2}\,\frac{\partial^2 S_{_{2-q}}}{\partial
V^2}-\left(\frac{\partial^2 S_{_{2-q}}}{\partial
E\,\partial V}\right)^2>{\cal A}_{_{2-q}}\,{\cal
B}_{_{2-q}} \ ,\label{tscq2}
\end{eqnarray}
where
\begin{equation}
{\cal A}_{_{2-q}}=\frac{q-1}{1+(q-1)\,S_{_{2-q}}} \
.\label{at}
\end{equation}
Eq. (\ref{at}) is a positive quantity for $q\in(1,\,2)$.
Consequently, it follows that both Eqs. (\ref{tscq1}) and
(\ref{tscq2}) are fulfilled if the concavity conditions are
satisfied. Differently, for $q\in(0,\,1)$ Eq. (\ref{at}) is
a negative quantity. For this range of values of $q$ the
thermodynamical stability of the equilibrium configuration
does not follow merely from the concavity conditions of
$S_{_{2-q}}$.\\ Such conclusion is in accordance with the
results discussed in Ref. \cite{Guerberoff}.

In spite of the success obtained by the Tsallis entropy in
the study of anomalous systems, others entropic forms, with
probability distribution function exhibiting an asymptotic
power law behavior, have been proposed by different
authors. Some of them belong to the family of the
Sharma-Taneja-Mittal entropy and we explore them in the
next examples.

\subsection{Abe entropy}

In Ref. \cite{Abe1} it has been presented a new entropy
containing the quantum group deformation structure, through
the requirement of the invariance under the interchange
$q\leftrightarrow q^{-1}$. This can be accomplished by
posing $r=\sqrt{1+\kappa^2}-1>0$ and
$q_{_A}=\sqrt{1+\kappa^2}+|\kappa|$, so that Eq.
(\ref{entropyst}) becomes
\begin{equation}
S_{q_{_{\rm A}}}=-\sum_{i=1}^W\frac{p_{_i}^{q_{_{\rm
A}}^{-1}}-p_{_i}^{q_{_{\rm A}}}}{q_{_{\rm A}}^{-1}-q_{_{\rm
A}}}=-\sum_{i=1}^Wp_{_i}\,\ln_{q_{_{\rm
A}}}\Big(p_{_i}\Big) \ ,\label{abeent1}
\end{equation}
with $1/2<q_{_A}\leq2$, and
\begin{equation}
\ln_{q_{_{\rm
A}}}(x)=\frac{x^{(q_{_A}^{-1})-1}-x^{q_{_A}-1}}{
q_{_A}^{-1}-q_{_A}} \ .\label{logabe}
\end{equation}
We remark that the inverse function of Eq. (\ref{logabe}),
namely $\exp_{q_{_{\rm A}}}(x)$, exists because Eq.
(\ref{logabe}) is a monotonic function, but its expression
cannot be given in term of elementary functions.

Entropy (\ref{abeent1}) has been applied in \cite{Abe7} to
the generalized statistical mechanics study of $q$-deformed
oscillators. The basic idea is to incorporate the
nonadditive feature of the energies of the systems having
the quantum group structures with generalized statistics.
It has been shown that for large value of $\partial
S_{q_{_{\rm A}}}/\partial E$ the deformation of the entropy
gives rise to significative deviations
of the Planck distribution with respect to the standard (undeformed) behavior.\\
In Ref. \cite{Abe1} it has been shown that Abe' entropy can
be expressed as a combination of Tsallis' entropy with
different deformation parameters. Consequently, many
physical proprieties of the former follows from the
physical proprieties of the latter. In particular, it can
be shown that it is Lesche stable \cite{Scarfone4}, and
fulfills the Pesin equality \cite{Lissia}.

In the microcanonical picture, the entropy (\ref{abeent1})
becomes
\begin{equation}
S_{q_{_{\rm A}}}=-\ln_{q_{_{\rm A}}}\left({1\over W}\right)
\ ,\label{abeent}
\end{equation}
and we introduce the function ${\mathcal I}_{q_{_A}}$,
through Eq. (\ref{i}), which assumes the expression
\begin{equation}
{\mathcal
I}_{q_{_A}}=\frac{1}{2}\left(W^{1-(q_{_A}^{-1})}+W^{1-q_{_A}}\right)
\ .\label{iaa}
\end{equation}
We recall that, according with Eq. (\ref{qmic}), Eq.
(\ref{iaa}) is a function of the entropy $S_{q_{_{\rm
A}}}$.\\ By taking into account the results of appendix A,
we have $1/2\leq{\mathcal I}_{q_{_{\rm A}}}\leq+\infty$,
depending on the value of $W$. After introducing the
threshold point through ${\mathcal I}_{q_{_A}}(W_{\rm
t})=1$, it follows that for $W_{\rm A}>W_{\rm t}$ and
$W_{\rm B}>W_{\rm t}$
\begin{equation}
S_{q_{_{\rm A}}}\left({\rm A}\cup{\rm
B}\right)>S_{q_{_A}}\left({\rm A}\right)+S_{q_{_A}}({\rm B}) \
.\label{aa1}
\end{equation}
In the same way, for $W_{\rm A}<W_{\rm t}$ and $W_{\rm B}<W_{\rm
t}$, we obtain
\begin{equation}
S_{q_{_{\rm A}}}\left({\rm A}\cup{\rm
B}\right)<S_{q_{_A}}\left({\rm A}\right)+S_{q_{_A}}({\rm B}) \
.\label{aa2}
\end{equation}
Temperature and pressure are given by
\begin{eqnarray}
&&{1\over T}={1\over{\mathcal I}_{q_{_A}}-\widetilde{q}_{_{\rm
A}}\,S_{q_{_{\rm A}}}}\,\frac{\partial\,S_{q_{_{\rm
A}}}}{\partial\,E} \ ,\\
&&P={T\over{\mathcal I}_{q_{_A}}-\widetilde{q}_{_{\rm
A}}\,S_{q_{_{\rm A}}}}\,\frac{\partial\,S_{q_{_{\rm
A}}}}{\partial\,V} \ ,
\end{eqnarray}
respectively, where $\widetilde{q}_{_{\rm A}}=(q_{_{\rm
A}}^{1/2}-q_{_{\rm A}}^{-1/2})^2/2$. They reduce to the
standard definition of temperature and pressure in the
$q_{_{\rm A}}\to 1$ limit.\\ The thermodynamic stability
conditions now read
\begin{eqnarray}
& &\frac{\partial^2S_{q_{_{\rm A}}}}{\partial E^2}<{\cal
A}_{q_{_{\rm A}}}\, \left(\frac{\partial S_{q_{_{\rm
A}}}}{\partial E}\right)^2 \ ,\label{tsca1}\\& & \nonumber\\
& &\frac{\partial^2 S_{q_{_{\rm A}}}}{\partial
E^2}\,\frac{\partial^2 S_{q_{_{\rm A}}}}{\partial
V^2}-\left(\frac{\partial^2 S_{q_{_{\rm A}}}}{\partial E\,\partial
V}\right)^2>{\cal A}_{q_{_{\rm A}}}\,{\cal B}_{q_{_{\rm A}}} \
,\label{tsca2}
\end{eqnarray}
with
\begin{equation}
{\cal A}_{q_{_{\rm A}}}=2\,\widetilde{q}_{_{\rm
A}}\,\frac{(\widetilde{q}_{_{\rm A}}+1)\,S_{q_{_{\rm
A}}}-{\mathcal I}_{q_{_A}}}{\left({\mathcal
I}_{q_{_A}}-\widetilde{q}_{_{\rm A}}\,S_{q_{_{\rm A}}}\right)^2} \
.\label{aaa}
\end{equation}
The sign of Eq. (\ref{aaa}) changes at the point
\begin{equation}
\widetilde W_{_t}=q_{_{\rm A}}^{2\,/(q_{_{\rm A}}-q_{_{\rm
A}}^{-1})} \ ,
\end{equation}
so that ${\cal A}_{q_{_{\rm A}}}<0$ for $W<\widetilde
W_{\rm t}$ and ${\cal A}_{q_{_{\rm A}}}>0$ for
$W>\widetilde W_{\rm t}$. On the other hand, observing that
$\widetilde W_{_t}<W_{\rm t}$, it follows that for
super-additive systems with $W_{\rm A}>W_{\rm t}$ and
$W_{\rm B}>W_{\rm t}$, the concavity conditions imply the
thermodynamic stability conditions.

It is worthy to observe that by posing
$r=1-\sqrt{1+\kappa^2}<0$ and
$q_{_A}=\sqrt{1+\kappa^2}-|\kappa|$ we obtain another
family of entropies embodies the symmetry
$q\leftrightarrow1/q$ given by
\begin{equation}
S_{q_{_{\rm
A}}}^\ast(W)=-\sum_{i=1}^W\frac{p_{_i}^{2-q_{_A}^{-1}}-p_{_i}^{2-q_{_A}}}{
q_{_A}-q_{_A}^{-1}}=-\sum_{i=1}^Wp_{_i}\,\ln^\ast_{q_{_{\rm
A}}}\Big(p_{_i}\Big) \ ,\label{abeentd1}
\end{equation}
with $1/2<q_{_A}\leq2$, where now
\begin{equation}
\ln_{q_{_{\rm
A}}}^\ast(x)=\frac{x^{1-q_{_A}^{-1}}-x^{1-q_{_A}}}{
q_{_A}-q_{_A}^{-1}} \ .\label{logabed}
\end{equation}
$\ln_{q_{_{\rm A}}}(x)$ and $\ln_{q_{_{\rm A}}}^\ast(x)$
are related in \cite{Naudts2}
\begin{equation}
\ln_{q_{_{\rm A}}}(x)=-\ln_{q_{_{\rm A}}}^\ast\left({1\over
x}\right) \ ,
\end{equation}
and the entropies (\ref{abeent1}) and (\ref{abeentd1}) are
dual
each other.\\
In the microcanonical picture Eq. (\ref{abeentd1}) becomes
\begin{equation}
S_{q_{_{\rm A}}}^\ast(W)=-\ln_{q_{_{\rm A}}}^\ast\left({1\over
W}\right) \ ,\label{abeentd}
\end{equation}
and because now the function ${\mathcal I}_{q_{_{\rm
A}}}^\ast>1$, the entropy (\ref{abeentd}) describes
super-additive systems: $S_{q_{_{\rm A}}}^\ast\left({\rm
A}\cup{\rm B}\right)>S_{q_{_A}}^\ast\left({\rm
A}\right)+S_{q_{_A}}^\ast({\rm B})$. The function ${\cal
A}_{q_{_{\rm A}}}^\ast>0$ and the concavity conditions for
the entropy (\ref{abeentd}) are enough to guarantee the
thermodynamic stability conditions of the system for any
values of the deformation parameter.

\subsection{Kaniadakis entropy}

As a last example, we discuss the entropic form introduced
previously in Ref. \cite{Kaniadakis3} which follows from
Eq. (\ref{entropyst}) after posed $r=0$:
\begin{equation}
S_{_{\kappa}}=-\sum_{i=1}^W\frac{p_{_i}^\kappa-p_{_i}^{-\kappa}}{2\,\kappa}=-\sum_{i=1}^W
p_{_i}\,\ln_{_{\{\kappa\}}}\Big(p_{_i}\Big) \
,\label{kent1}
\end{equation}
where $|\kappa|<1$, and the $\kappa$-logarithm
$\ln_{_{\{\kappa\}}}(x)$ is defined by
\begin{equation}
\ln_{_{\{\kappa\}}}(x)=\frac{x^\kappa-x^{-\kappa}}{2\,\kappa}
\ .\label{klog}
\end{equation}
Its inverse function, the $\kappa$-exponential, is given by
\begin{equation}
\exp_{_{\{\kappa\}}}(x)=\left(\sqrt{1+\kappa^2\,x^2}
+\kappa\,x\right)^{1/\kappa} \ ,
\end{equation}
and satisfies the relation
\begin{equation}
\exp_{_{\{\kappa\}}}(x)\,\exp_{_{\{\kappa\}}}(-x)=1 \ ,
\end{equation}
which means that it increases for $x\to\infty$ and
decreases
for $x\to-\infty$ with the same steepness.\\
Remarkably, the $\kappa$-logarithm and the
$\kappa$-exponential fulfill the two following mathematical
proprieties
\begin{eqnarray}
&&\exp_{_{\{\kappa\}}}(x)\,\exp_{_{\{\kappa\}}}(y)=\exp_{_{\{\kappa\}}}(x\stackrel{\kappa}{\oplus}
y)
\ ,\label{ke}\\
&&\ln_{_{\{\kappa\}}}(x\,y)=\ln_{_{\{\kappa\}}}(x)\stackrel{\kappa}{\oplus}\ln_{_{\{\kappa\}}}(y)
\ ,\label{kl}
\end{eqnarray}
where the $\kappa$-deformed sum is defined in
\cite{Scarfone3}
\begin{equation}
x\stackrel{\kappa}{\oplus}y=x\,\sqrt{1+\kappa^2\,y^2}+y\,\sqrt{1+\kappa^2\,x^2}
\ .\label{ksum}
\end{equation}
Eqs. (\ref{ke}) and (\ref{kl}) reduce, in the $\kappa\to0$
limit, to the well known relations
$\exp(x)\,\exp(y)=\exp(x+y)$ and $\ln(x\,y)=\ln(x)+\ln(y)$,
respectively,
according to $x\stackrel{0}{\oplus}y=x+y$.\\
In Ref. \cite{Kaniadakis2} it has been shown that the
$\kappa$-sum emerges naturally within the Einstein special
relativity. In fact, following the same argument presented
in \cite{Kaniadakis2} it is possible to link the
relativistic sum of the velocities
\begin{equation}
v_1\oplus^cv_2=\frac{v_1+v_2}{1+v_1\,v_2/c^2} \ ,
\end{equation}
with the $\kappa$-sum (\ref{ksum}) in the sense of
\begin{equation}
p(v_1)\stackrel{\kappa}{\oplus}p(v_2)=p(v_1\oplus^cv_2) \ ,
\end{equation}
where $\kappa=1/m\,c$ and $p(v)=m\,v/\sqrt{1-v^2/c^2}$ is
the relativistic momentum of a particle with rest mass $m$.
In this way, the origin of the $\kappa$-deformation are
related to the finite value of light speed $c$. In
particular, in the classical limit $c\to\infty$ the
parameter $\kappa$ approaches to zero and the
$\kappa$-entropy reduces to the
Boltzmann-Gibbs one.\\
In agreement with this interpretation, we consider
statistical systems (physical or not) that can achieve an
equilibrium configuration through an exchange of
information between the parts of the system, propagating
with a limiting velocity, like the light speed in the
relativity theory. For these systems it is reasonable
suppose that a mechanism similar to the one describe above,
in the framework of the special relativity, can arise, so
that the $\kappa$-deformation can occurs. In this way, the
$\kappa$-entropy can be successfully applied in the study
of the statistical proprieties of these systems.\\
An important physical example where the
$\kappa$-distribution has been successfully applied is in
the reproduction of the energy distribution of the fluxes
of cosmic rays \cite{Kaniadakis2} (see also
\cite{Tsallisrays}). Moreover the $\kappa$-entropy has been
applied in the study of the fracture propagation in brittle
material, showing a good accordance with the results
obtained experimentally and with the ones obtained through
numerical simulations \cite{Scarfonefrac}.\\ Finally, we
recall that, like the previous one-parameter deformed
entropic forms, also the $\kappa$-entropy fulfils many
physically relevant proprieties. In particular, in Ref.
\cite{Scarfone5}, it has been shown its experimental
stability whereas in Ref. \cite{Lissia} the finite entropy
production in time unit in connection with the Pesin
identity for a system described by the entropy
(\ref{kent1}) has been discussed.

In the microcanonical picture, Eq. (\ref{kent1}) becomes
\begin{equation}
S_{_{\kappa}}=\ln_{_{\{\kappa\}}}(W) \ ,\label{kent}
\end{equation}
and we introduce the function ${\mathcal I}_{_{\kappa}}$
that, according to Eq. (\ref{qmic}), can be written in
\begin{equation}
{\mathcal I}_{_{\kappa}}=\sqrt{1+\kappa^2\,S_{_{\kappa}}^2} \
,\label{ik}
\end{equation}
so that ${\mathcal I}_{_{\kappa}}\geq1$. As a consequence
Eq. (\ref{comp1}) becomes
\begin{eqnarray}
\nonumber S_{_{\kappa}}\left({\rm A}\cup{\rm
B}\right)&=&S_{_{\kappa}}\left({\rm
A}\right)\,\sqrt{1+\kappa^2\,\left[S_{_{\kappa}}\left({\rm
B}\right)\right]^2}\\&+&S_{_{\kappa}}\left({\rm
B}\right)\,\sqrt{1+\kappa^2\,\left[S_{_{\kappa}}\left({\rm
A}\right)\right]^2} \ ,\label{ksum1}
\end{eqnarray}
which can be also written in
\begin{equation}
S_{_{\kappa}}\left({\rm A}\cup{\rm
B}\right)=S_{_{\kappa}}\left({\rm
A}\right)\stackrel{\kappa}{\oplus}S_{_{\kappa}}\left({\rm
B}\right) \ ,\label{ksum2}
\end{equation}
according to Eq. (\ref{ksum}). Eqs. (\ref{ksum1}) or
(\ref{ksum2}) imply the relation
\begin{equation}
S_{_{\kappa}}\left({\rm A}\cup{\rm
B}\right)>S_{_{\kappa}}\left({\rm
A}\right)+S_{_{\kappa}}\left({\rm B}\right) \ ,\label{k-sum}
\end{equation}
stating that the $\kappa$-entropy in the microcanonical picture is always super-additive.\\
Temperature and pressure are given by
\begin{equation}
{1\over
T}={1\over\sqrt{1+\kappa^2\,S_{_\kappa}^2}}\,\frac{\partial\,S_{_{\kappa}}}{\partial
E} \ ,
\end{equation}
\begin{equation}
P={T\over\sqrt{1+\kappa^2\,S_{_\kappa}^2}}\,\frac{\partial\,S_{_{\kappa}}}{\partial
V} \ ,\label{tpk}
\end{equation}
respectively, and reduce to the standard definitions in the $\kappa\to 0$ limit.\\
Finally, the thermodynamic stability conditions become
\cite{Wada2,Wada3}
\begin{eqnarray}
& &\frac{\partial^2S_{_\kappa}}{\partial E^2}< {\cal
A}_{_\kappa}\,\left(\frac{\partial S_{_\kappa}}{\partial
E}\right)^2 \ ,\label{tsck1}\\
& &\\ \nonumber & &\frac{\partial^2
S_{_\kappa}}{\partial E^2}\,\frac{\partial^2 S_{_\kappa}}{\partial
V^2}-\left(\frac{\partial^2 S_{_\kappa}}{\partial E\,\partial
V}\right)^2>{\cal A}_{_\kappa}\,{\cal B}_{_\kappa} \
,\label{tsck2}
\end{eqnarray}
where
\begin{equation}
{\cal A}_{_\kappa}=\frac{\kappa^2
S_{_\kappa}}{1+\kappa^2\,S_{_\kappa}^2} \ .\label{ak}
\end{equation}
The function (\ref{ak}) is always positive and, as a
consequence, the concavity conditions for the entropy
(\ref{kent}) are enough to guarantee the thermodynamic
stability conditions of the system for any values of the
deformation parameter.

\sect{Concluding remarks} In the present work we have
investigated the thermodynamic stability conditions for a
microcanonical system described by the Sharma-Taneja-Mittal
entropy, and their relation with the
concavity conditions for this entropy.\\
The main results can be summarized in the following two points:\\
Firstly, we have analyzed the ``composability'' rule for
statistically independent systems described by the entropy
(\ref{Boltzmann}). It has been shown that the parameter space
$\cal R$ can be split into two disjoint regions. In the region
${\cal R}\big|_{r\leq0}$ the entropy $S_{_{\kappa,\,r}}$ shows a
super-additivity behavior: $S_{_{\kappa,\,r}}({\rm A}\cup{\rm
B})>S_{_{\kappa,\,r}}({\rm A})+S_{_{\kappa,\,r}}({\rm B})$.
Otherwise, in the region ${\cal R}\big|_{r>0}$ the behavior of the
entropy is not well defined, depending on the {\em size} of the
two systems A and B. In particular it has been shown that, given
the threshold point $W_{\rm t}(\kappa,\,r)>1$, when the {\em size}
of the two parts A and B are smaller than $W_{\rm t}$, in the
sense of $W_{\rm A}<W_{\rm t}$ and $W_{\rm B}<W_{\rm t}$, the
system exhibits a sub-additive behavior $S_{_{\kappa,\,r}}({\rm
A}\cup{\rm B})<S_{_{\kappa,\,r}}({\rm A})+S_{_{\kappa,\,r}}({\rm
B})$, becoming super-additive when both $W_{\rm
A}>W_{\rm t}$ and $W_{\rm B}>W_{\rm t}$.\\
Secondly, we have inquired on the thermodynamic stability
conditions of the equilibrium configuration. In the
Boltzmann theory the concavity conditions imply the
thermodynamic stability conditions. Such a situation
changes when the system is described by the entropy
$S_{_{\kappa,\,r}}$. We have shown that, starting from an
equilibrium configuration of the system A$\cup$B, and
supposed an exchange of a small quantity of heat and/or
work between the two parts A and B, assumed statistically
independent, if the entropy of the system
$S_{_{\kappa,\,r}}({\rm A}\cup{\rm B})$ is larger than the
sum of the entropy of the two systems
$S_{_{\kappa,\,r}}({\rm A})$ and $S_{_{\kappa,\,r}}({\rm
B})$, the concavity conditions still imply the
thermodynamic stability conditions. In the opposite
situation, in spite of the concavity of
$S_{_{\kappa,\,r}}$, stability requires {\em large}
systems, in the sense of $W_{\rm A}>W_{\rm t}$ and $W_{\rm
B}>W_{\rm t}$.

\app \sect{} In this Appendix we summarize some mathematical
properties of the deformed logarithm \cite{Scarfone1,Scarfone2}
\begin{equation}
\ln_{_{\{\kappa,\,r\}}}(x)=\frac{x^{r+\kappa}-
x^{r-\kappa}}{2\,\kappa} \ .\label{log}
\end{equation}
Let $\mathcal R$ be the region in the parametric space, defined by
\begin{equation}
I\!\!R^2\supset{\mathcal R}=\left\{
\begin{array}{l}
\hspace{4mm}-|\kappa|\leq r\leq|\kappa|\hspace{12mm} {\rm if} \
0\leq|\kappa|<{1\over2} \ ,\\
|\kappa|-1<r<1-|\kappa|\hspace{5mm} {\rm if} \
{1\over2}\leq|\kappa|<1 \ .
\end{array}
\right.\label{quadri}
\end{equation}
For any $(\kappa,\,r)\in{\mathcal R}$, the
$\ln_{_{\{\kappa,\,r\}}}(x)=\ln_{_{\{-\kappa,\,r\}}}(x)$, is a
continuous, monotonic, increasing and concave function for $x\in
I\!\!R^+$, with $\ln_{_{\{\kappa,\,r\}}}(I\!\!R^+)\subseteq
I\!\!R$, fulfilling the relation
$\int_0^1\ln_{_{\{\kappa,\,r\}}}(x^{\pm1})\,dx=\mp1/[(1\pm
r)^2-\kappa^2]$. The standard logarithm is recovered in the
$(\kappa,\,r)\to(0,\,0)$ limit: $\ln_{_{\{0,\,0\}}}(x)=\ln(x)$.\\
Eq. (\ref{log}) satisfies the relation
\begin{equation}
\ln_{_{\{\kappa,\,r\}}}(x)=-\ln_{_{\{\kappa,\,-r\}}}
\left(\frac{1}{x}\right) \ ,\label{dual}
\end{equation}
and for $r=0$, reproduces the well known properties of the
standard logarithm
$\ln_{_{\{\kappa,\,0\}}}(x)=-\ln_{_{\{\kappa,\,0\}}}(1/x)$. From
Eq. (\ref{dual}) it follows that the behavior of
$\ln_{_{\{\kappa,\,r\}}}(x)$ for $r>0$ and $0\leq x\leq1$ is
related to the one for $r<0$ and $x>1$.\\
We observe that the deformed logarithm is a solution of the
following differential-functional equation
\begin{equation}
\frac{d}{d\,x}
\left[x\,\ln_{_{\{\kappa,\,r\}}}(x)\right]=\lambda\,
\ln_{_{\{\kappa,\,r\}}}\left({x\over\alpha}\right)
\ ,
\end{equation}
with the following boundary conditions
$\ln_{_{\{\kappa,\,r\}}}(1)=0$ and
$d\,\ln_{_{\{\kappa,\,r\}}}(x)/d\,x\Big|_{x=1}=1$ and the
constants $\alpha$ and $\lambda$ given in Eqs.
(\ref{alpha})-(\ref{lambda}).

The inverse function of Eq. (\ref{log}), namely the
deformed exponential $\exp_{_{\{\kappa,\,r\}}}(x)$, exists
for any $(\kappa,\,r)\in{\mathcal R}$, since
$\ln_{_{\{\kappa,\,r\}}}(x)$ is a strictly monotonic
function. Its analytical properties are well-defined and
follow from the corresponding ones of the deformed
logarithm. Nevertheless, the explicit expression of the
$\exp_{_{\{{\scriptstyle \kappa,\,r}\}}}(x)$ can be
obtained only for particular relationships
between $r$ and $\kappa$.\\
By inspection there exists a point $x_{_0}(\kappa,\,r)$ such that
the inequality $\ln_{_{\{\kappa,\,r\}}}(x)\geq\ln x$ holds for $x>
x_{_0}(\kappa,\,r)$. We see that $x_{_0}(\kappa,\,r)$ is a
monotonic decreasing function w.r.t $r$ with
$x_{_0}(\kappa,\,r)>1$ for $r<0$ and $x_{_0}(\kappa,\,r)<1$ for
$r>0$. In particular $x_{_0}(\kappa,\,-\kappa)=+\infty$,
$x_{_0}(\kappa,\,0)=1$
and $x_{_0}(\kappa,\,\kappa)=0$.\\
Entropy $S_{_{\kappa,r}}(p)$, introduced in Eq.
(\ref{entropyst}), is related to the function
$\ln_{_{\{\kappa,\,r\}}}(x)$ through the relation
\begin{equation}
S_{_{\kappa,r}}(p)=-\sum_{i=1}^Wp_{_i}\,\ln_{_{\{\kappa,\,r\}}}(p_{_i})
\ .\label{sl}
\end{equation}

The deformed logarithm (\ref{log}) can be obtained from the
two-parametric generalization of the Jackson derivative,
previously proposed in Ref. \cite{Chakrabarti}
\begin{equation}
\frac{d_{_{\kappa,\,r}}\,f(x)}{d_{_{\kappa,\,r}}\,x}
=\frac{f((r+\kappa)\,x)-f((r-\kappa)\,x)}{2\,\kappa\,x} \
,\label{Jackson}
\end{equation}
by posing
\begin{equation}
\ln_{_{\{\kappa,\,r\}}}(x)=\frac{d_{_{\kappa,\,r}}\,x^y}
{d_{_{\kappa,\,r}}\,y}\Bigg|_{y=1} \ .
\end{equation}
Some properties of the deformed logarithm can be naturally
understood as those of the generalized Jackson derivative
(\ref{Jackson}). For instance, from the generalized Leibnitz rule
\begin{eqnarray}
\nonumber
\frac{d_{_{\kappa,\,r}}\,f(x)\,g(x)}{d_{_{\kappa,\,r}}\,x}&=&
\frac{d_{_{\kappa,\,r}}\,f(x)}{d_{_{\kappa,\,r}}\,x}
\,g((r+\kappa)\,x)\\&+&
f((r-\kappa)\,x)\,\frac{d_{_{\kappa,\,r}}\,g(x)}
{d_{_{\kappa,\,r}}\,x} \ ,
\end{eqnarray}
we obtain the following useful relation
\begin{equation}
\ln_{_{\{\kappa,\,r\}}}(x\,y)=x^{r+\kappa}
\,\ln_{_{\{\kappa,\,r\}}}
(y)+y^{r-\kappa}\,\ln_{_{\{\kappa,\,r\}}}(x) \ ,\label{rl1}
\end{equation}
and by using the identity
$y^{r-\kappa}=y^{r+\kappa}-2\,\kappa\,\ln_{_{\{\kappa,\,r\}}}(y)$,
Eq. (\ref{rl1}) becomes
\begin{eqnarray}
\nonumber \ln_{_{\{\kappa,\,r\}}}(x\,y)&=&x^{r+\kappa}
\,\ln_{_{\{\kappa,\,r\}}}
(y)+y^{r+\kappa}\,\ln_{_{\{\kappa,\,r\}}}(x)\\
&-&2\,\kappa\,\ln_{_{\{\kappa,\,r\}}}(x)\,\ln_{_{\{\kappa,\,r\}}}(y)
\ .\label{rl2}
\end{eqnarray}
Moreover, recalling the $\kappa\leftrightarrow-\kappa$ symmetry,
Eq. (\ref{rl2}) can be rewritten in
\begin{eqnarray}
\nonumber
\ln_{_{\{\kappa,\,r\}}}(x\,y)&=&u_{_{\{\kappa,\,r\}}}(x)\,\ln_{_{\{\kappa,\,r\}}}
(y)\\&+&u_{_{\{\kappa,\,r\}}}(x)\,\ln_{_{\{\kappa,\,r\}}}(y)
\ ,\label{rl3}
\end{eqnarray}
where we have introduced the function
\begin{equation}
u_{_{\{\kappa,\,r\}}}(x)={x^{r+\kappa}+x^{r-\kappa}\over2} \
.\label{u}
\end{equation}
For any $(\kappa,\,r)\in{\cal R}$ the function
$u_{_{\{\kappa,\,r\}}}(x)=u_{_{\{-\kappa,\,r\}}}(x)$, is
continuous for $x\in I\!\!R^+$, with
$u_{_{\{\kappa,\,r\}}}(I\!\!R^+)\subseteq I\!\!R^+$,
$u_{_{\{\kappa,\,r\}}}(0)=u_{_{\{\kappa,\,r\}}}(+\infty)=+\infty$
for $r\not=|\kappa|$, satisfies the relation
$u_{_{\{\kappa,\,r\}}}(x)=u_{_{\{\kappa,\,-r\}}}(1/x)$ and reduces
to the unity in the $(\kappa,\,r)\to(0,\,0)$ limit:
$u_{_{\{0,\,0\}}}(x)=1$. It reaches the minimum values
\begin{equation}
u_{_{\{\kappa,\,r\}}}(x_{_m})=\kappa\,\frac{\left(\kappa-r\right)}{\mbox{\raisebox{-1mm}
{$\left(\kappa+r\right)$}}}
^{\scriptscriptstyle{(r-\kappa)/2\,\kappa}}_
{\mbox{\raisebox{3.5mm}
{$\scriptscriptstyle{(r+\kappa)/2\,\kappa}$}}}
 \ ,\label{um}
\end{equation}
at
\begin{equation}
x_{_m}=\left(\frac{\kappa-r}{\kappa+r}\right)^{1/2\,\kappa} \
.\label{xm}
\end{equation}
In particular, for any $(\kappa,\,r)\in{\cal R}\big|_{r\leq0}$,
$x_{_m}\geq1$, and taking into account that
$u_{_{\{\kappa,\,r\}}}(1)=1$, it follows $1\leq
u_{_{\{\kappa,\,r\}}}(x)\leq\infty$ when $x\in(0,\,1)$. For any
$(\kappa,\,r)\in{\cal R}\big|_{r>0}$, from Eq. (\ref{xm}) we
obtain $0\leq x_{_m}\leq1$ and from Eq. (\ref{um}) it follows
$1/2\leq u_{_{\{\kappa,\,r\}}}(x_{_m})\leq1$.\\ By inspection, it
follows that there exist a threshold point $x_{_t}(\kappa,\,r)$,
defined by $u_{_{\{\kappa,\,r\}}}(x_{_t})=1$. $x_{_t}(\kappa,\,r)$
is monotonic decreasing function w.r.t. $r$, with
$x_{_t}(\kappa,\,-\kappa)=+\infty$, $x_{_t}(\kappa,\,0)=1$ and
$x_{_t}(\kappa,\,\kappa)=0$, such that $1/2\leq
u_{_{\{\kappa,\,r\}}}(x_{_m})\leq u_{_{\{\kappa,\,r\}}}(x)\leq1$
for $x\geq x_{_t}(\kappa,\,r)$ and $1\leq
u_{_{\{\kappa,\,r\}}}(x)\leq+\infty$ for $0\leq x\leq
x_{_t}(\kappa,\,r)$.\\
Finally, we remark that the function (\ref{u}) fulfills the
following relation
\begin{equation}
u_{_{\{\kappa,\,r\}}}(x\,y)=u_{_{\{\kappa,\,r\}}}(x)\,u_{_{\{\kappa,\,r\}}}(y)+\kappa^2\,
\ln_{_{\{\kappa,\,r\}}} (x)\,\ln_{_{\{\kappa,\,r\}}} (y) \
.\label{sumu}
\end{equation}
We observe that, like the deformed logarithm, the function
$u_{_{\{\kappa,\,r\}}}(x)$ is a solution of the
differential-functional equation (\ref{condint}) with the
following boundary conditions $u_{_{\{\kappa,\,r\}}}(1)=1$
and $d\,u_{_{\{\kappa,\,r\}}}(x)/d\,x\Big|_{x=1}=r$ and the
constants $\alpha$
and $\lambda$ given in Eqs. (\ref{alpha})-(\ref{lambda}). Moreover the two functions
$\ln_{_{\{\kappa,\,r\}}}(x)$ and $u_{_{\{\kappa,\,r\}}}(x)$ are related by the relation
\begin{equation}
u_{_{\{\kappa,\,r\}}}(x)=x^{r+\kappa}-\kappa\,\ln_{_{\{\kappa,\,r\}}}(x)
\ .
\end{equation}
The function ${\cal I}_{_{\kappa,\,r}}(p)$, introduced in
Eq. (\ref{i}), is related to the function
$u_{_{\{\kappa,\,r\}}}(x)$ through the relation
\begin{equation}
{\cal
I}_{_{\kappa,\,r}}(p)=\sum_{i=1}^Wp_{_i}\,u_{_{\{\kappa,\,r\}}}(p_{_i})
\ .\label{ii}
\end{equation}
From the definitions (\ref{sl}), (\ref{ii}) and Eqs. (\ref{rl3}),
(\ref{sumu}), we obtain the useful relations
\begin{eqnarray}
\nonumber &&S_{_{\kappa,\,r}}({\rm A}\cup{\rm
B})=S_{_{\kappa,\,r}}({\rm A})\,{\mathcal
I}_{_{\kappa,\,r}}({\rm B})+{\mathcal
I}_{_{\kappa,\,r}}({\rm A})\,S_{_{\kappa,\,r}}({\rm
B}) \ ,\\\label{s1}\\
\nonumber&&{\mathcal I}_{_{\kappa,\,r}}({\rm A}\cup{\rm
B})={\mathcal I}_{_{\kappa,\,r}}({\rm A})\,{\mathcal
I}_{_{\kappa,\,r}}({\rm
B})+\kappa^2\,S_{_{\kappa,\,r}}({\rm
A})\,S_{_{\kappa,\,r}}({\rm B}) \ ,\\\label{s2}
\end{eqnarray}
stating the additivity of $S_{_{\kappa,r}}$ and ${\mathcal
I}_{_{\kappa,\,r}}$ for statistical independent systems $p^{{\rm
A}\cup{\rm B}}=\{p_{_i}^{\rm A}\,p_{_j}^{\rm B}\}$ with
$i=1,\,\cdots,\,W_{\rm A}$ and $j=1,\,\cdots,\,W_{\rm B}$.
\sect{}

In this Appendix we derive the equilibrium conditions given in
Eqs. (\ref{0law})-(\ref{mech}), and the
thermodynamic stability conditions given in Eqs. (\ref{tsc2})-(\ref{tsc3}).\\
Let us suppose that the whole system A$\cup$B initially at
equilibrium undergoes a small transfer of heat and/or work between
the two parts A and B, with the constraints
\begin{eqnarray}
&&\delta(E_{\rm A}+E_{\rm B})=0 \ ,\label{b1}\\
&&\delta(V_{\rm A}+V_{\rm B})=0 \ .
\end{eqnarray}
Recalling that the entropy evaluated at an equilibrium
configuration is a maximum, we have
\begin{equation}
S_{_{\kappa,\,r}}({\rm A}\cup{\rm B})> S_{_{\kappa,\,r}}(({\rm
A}+\delta{\rm A})\cup({\rm B}+\delta{\rm B})) \ ,\label{bb}
\end{equation}
and taking into account of Eq. (\ref{s1}) it follows
\begin{eqnarray}
\nonumber & &S_{_{\kappa,\,r}}({\rm A})\,{\mathcal
I}_{_{\kappa,\,r}}({\rm B})+{\mathcal
I}_{_{\kappa,\,r}}({\rm A})\,S_{_{\kappa,\,r}}({\rm B})>\\
\nonumber& &S_{_{\kappa,\,r}}({\rm A}+\delta{\rm
A})\,{\mathcal I}_{_{\kappa,\,r}}({\rm B}-\delta{\rm
B})\\&+&{\mathcal I}_{_{\kappa,\,r}}({\rm A}+\delta{\rm
A})\,S_{_{\kappa,\,r}}({\rm B}-\delta{\rm B}) \ .\label{b2}
\end{eqnarray}
We expand the r.h.s of Eq. (\ref{b2}) up to the second order in
$\delta\,E$ and $\delta\,V$, where $\delta E\equiv \delta E_{\rm
A}=-\delta E_{\rm B}$ and $\delta V\equiv \delta V_{\rm A}=-\delta
V_{\rm B}$. According to the MaxEnt principle, the first order
terms must vanish
\begin{eqnarray}
\nonumber&&
\hspace{-5mm}\left(\frac{\partial\,S_{_{\kappa,\,r}}({\rm
A})}{\partial\,E_{\rm A}}\,{\mathcal
I}_{_{\kappa,\,r}}({\rm B})+\frac{\partial\,{\mathcal
I}_{_{\kappa,\,r}}({\rm A})}{\partial\,E_{\rm
A}}\,S_{_{\kappa,\,r}}({\rm B})\right.\\
\nonumber&&\left.-S_{_{\kappa,\,r}}({\rm
A})\,\frac{\partial\,{\mathcal I}_{_{\kappa,\,r}}({\rm
B})}{\partial\,E_{\rm B}}-{\mathcal I}_{_{\kappa,\,r}}({\rm
A})\,\frac{\partial\,S_{_{\kappa,\,r}}({\rm
B})}{\partial\,E_{\rm
B}}\right)\,\delta E\\
\nonumber
&&\hspace{-5mm}+\left(\frac{\partial\,S_{_{\kappa,\,r}}({\rm
A})}{\partial\,V_{\rm A}}\,{\mathcal
I}_{_{\kappa,\,r}}({\rm B})+\frac{\partial\,{\mathcal
I}_{_{\kappa,\,r}}({\rm A})}{\partial\,V_{\rm
A}}\,S_{_{\kappa,\,r}}({\rm B})\right.\\
\nonumber&&\left.-S_{_{\kappa,\,r}}({\rm
A})\,\frac{\partial\,{\mathcal I}_{_{\kappa,\,r}}({\rm
B})}{\partial\,V_{\rm B}}-{\mathcal I}_{_{\kappa,\,r}}({\rm
A})\,\frac{\partial\,S_{_{\kappa,\,r}}({\rm
B})}{\partial\,V_{\rm
B}}\right)\,\delta V=0 \ .\\
\label{b3}
\end{eqnarray}
By using the relations
\begin{eqnarray}
\frac{\partial\,{\mathcal I}_{_{\kappa,\,r}}}{\partial
X}=\frac{\kappa^2\,S_{_{\kappa,\,r}}-r\,{\mathcal
I}_{_{\kappa,\,r}}}{{\mathcal
I}_{_{\kappa,\,r}}-r\,S_{_{\kappa,\,r}}}\,\frac{\partial
S_{_{\kappa,\,r}}}{\partial X} \ ,
\end{eqnarray}
with $X\equiv E$ or $X\equiv V$, Eq. (\ref{b3}) becomes
\begin{eqnarray}
\nonumber& &\left({\mathcal
I}_{_{\kappa,\,r}}-r\,S_{_{\kappa,\,r}}\right)\Bigg|_{{\rm
A}\cup{\rm B}}\left[{1\over{\mathcal
I}_{_{\kappa,\,r}}-r\,S_{_{\kappa,\,r}}}\,\frac{\partial
S_{_{\kappa,\,r}}}{\partial E}\Bigg|_{\rm A}\,\delta E\right.\\
\nonumber&&+{1\over{\mathcal
I}_{_{\kappa,\,r}}-r\,S_{_{\kappa,\,r}}}\,\frac{\partial
S_{_{\kappa,\,r}}}{\partial V}\Bigg|_{\rm A}\,\delta V
-{1\over{\mathcal
I}_{_{\kappa,\,r}}-r\,S_{_{\kappa,\,r}}}\,\frac{\partial
S_{_{\kappa,\,r}}}{\partial E}\Bigg|_{\rm B}\,\delta E\\
&&\left.-{1\over{\mathcal
I}_{_{\kappa,\,r}}-r\,S_{_{\kappa,\,r}}}\,\frac{\partial
S_{_{\kappa,\,r}}}{\partial V}\Bigg|_{\rm B}\,\delta
V\right]=0 \ ,\label{b6}
\end{eqnarray}
where we use the relations (\ref{s1}) and (\ref{s2}).\\ Taking
into account that
\begin{eqnarray}
& &{\mathcal I}_{_{\kappa,\,r}}\left({1\over
W}\right)-r\,S_{_{\kappa,\,r}}\left({1\over
W}\right)\\&=&{1\over W}\,\frac{d}{d\,\left(1/
W\right)}\,\ln_{_{\{\kappa,\,r\}}}\left({1\over W}\right)>0
\ ,\label{b14}
\end{eqnarray}
through Eq. (\ref{b6}) the two equilibrium conditions follow
\begin{eqnarray}
\nonumber &&{1\over{\mathcal
I}_{_{\kappa,\,r}}-r\,S_{_{\kappa,\,r}}}
\,{\partial\,S_{_{\kappa,\,r}}\over\partial\,E}\Bigg|_{\rm
A}={1\over{\mathcal
I}_{_{\kappa,\,r}}-r\,S_{_{\kappa,\,r}}}\,
{\partial\,S_{_{\kappa,\,r}}\over\partial\,E}\Bigg|_{\rm B} \ ,\\
&&\\ \nonumber &&{1\over{\mathcal
I}_{_{\kappa,\,r}}-r\,S_{_{\kappa,\,r}}}\,{\partial\,S_{_{\kappa,\,r}}
\over\partial\,V}\Bigg|_{\rm A}={1\over{\mathcal
I}_{_{\kappa,\,r}}-r\,S_{_{\kappa,\,r}}}\,
{\partial\,S_{_{\kappa,\,r}}\over\partial\,V}\Bigg|_{\rm B}
\ ,\\
\end{eqnarray}
which coincide with Eqs. (\ref{0law}) and (\ref{mech}).\\

In order to obtain the thermodynamic stability conditions
let us consider the second order terms in the expansion of
Eq. (\ref{b2})
\begin{eqnarray}
\nonumber & &{1\over2}\left[{\mathcal I}_{_{\kappa,\,r}}({\rm
A})\,\frac{\partial^2 S_{_{\kappa,\,r}}({\rm B})}{\partial E_{\rm
B}^2}-2\,\frac{\partial{\mathcal I}_{_{\kappa,\,r}}({\rm
A})}{\partial E_{\rm A}}\,\frac{S_{_{\kappa,\,r}}({\rm
B})}{\partial E_{\rm B}}\right.\\
\nonumber &+&\left.\!\!\!\frac{\partial^2 {\mathcal
I}_{_{\kappa,\,r}}({\rm A})}{\partial E_{\rm
A}^2}\,S_{_{\kappa,\,r}}({\rm B})+{\mathcal
I}_{_{\kappa,\,r}}({\rm B})\,\frac{\partial^2
S_{_{\kappa,\,r}}({\rm A})}{\partial E_{\rm
A}^2}\right.\\
\nonumber &-&\!\!\!\left.2\,\frac{\partial{\mathcal
I}_{_{\kappa,\,r}}({\rm B})}{\partial E_{\rm
B}}\,\frac{S_{_{\kappa,\,r}}({\rm A})}{\partial E_{\rm
A}}+\frac{\partial^2 {\mathcal I}_{_{\kappa,\,r}}({\rm
B})}{\partial E_{\rm
B}^2}\,S_{_{\kappa,\,r}}({\rm A})\right]\,(\delta E)^2\\
\nonumber &+&\!\!\!{1\over2}\left[{\mathcal
I}_{_{\kappa,\,r}}({\rm A})\,\frac{\partial^2
S_{_{\kappa,\,r}}({\rm B})}{\partial V_{\rm
B}^2}-2\,\frac{\partial{\mathcal I}_{_{\kappa,\,r}}({\rm
A})}{\partial V_{\rm A}}\,\frac{S_{_{\kappa,\,r}}({\rm
B})}{\partial V_{\rm B}}\right.\\
\nonumber &+&\!\!\!\left.\frac{\partial^2 {\mathcal
I}_{_{\kappa,\,r}}({\rm A})}{\partial V_{\rm
A}^2}\,S_{_{\kappa,\,r}}({\rm B})+{\mathcal
I}_{_{\kappa,\,r}}({\rm B})\,\frac{\partial^2
S_{_{\kappa,\,r}}({\rm A})}{\partial V_{\rm
A}^2}\right.\\
\nonumber &-&\!\!\!\left.2\,\frac{\partial{\mathcal
I}_{_{\kappa,\,r}}({\rm B})}{\partial V_{\rm
B}}\,\frac{S_{_{\kappa,\,r}}({\rm A})}{\partial V_{\rm
A}}+\frac{\partial^2 {\mathcal I}_{_{\kappa,\,r}}({\rm
B})}{\partial V_{\rm
B}^2}\,S_{_{\kappa,\,r}}({\rm A})\right]\,(\delta V)^2\\
\nonumber &+&\!\!\!\left[{\mathcal I}_{_{\kappa,\,r}}({\rm
A})\,\frac{\partial^2 S_{_{\kappa,\,r}}({\rm B})}{\partial E_{\rm
B}\,\partial V_{\rm B}}-\frac{\partial{\mathcal
I}_{_{\kappa,\,r}}({\rm A})}{\partial V_{\rm A}}\,\frac{\partial
S_{_{\kappa,\,r}}({\rm B})}{\partial E_{\rm
B}}\right.\\
\nonumber &-&\!\!\!\left.\,\frac{\partial{\mathcal
I}_{_{\kappa,\,r}}({\rm A})}{\partial E_{\rm A}}\,\frac{\partial
S_{_{\kappa,\,r}}({\rm B})}{\partial V_{\rm B}}+\frac{\partial^2
{\mathcal I}_{_{\kappa,\,r}}({\rm A})}{\partial E_{\rm
A}\,\partial V_{\rm A}}\,S_{_{\kappa,\,r}}({\rm B})\right.\\
\nonumber &+&\!\!\!\left.{\mathcal I}_{_{\kappa,\,r}}({\rm
B})\,\frac{\partial^2 S_{_{\kappa,\,r}}({\rm A})}{\partial E_{\rm
A}\,\partial V_{\rm A}}-\frac{\partial{\mathcal
I}_{_{\kappa,\,r}}({\rm B})}{\partial V_{\rm B}}\,\frac{\partial
S_{_{\kappa,\,r}}({\rm A})}{\partial E_{\rm
A}}\right.\\
\nonumber &-&\!\!\!\left.\frac{\partial{\mathcal
I}_{_{\kappa,\,r}}({\rm B})}{\partial E_{\rm
B}}\,\frac{\partial S_{_{\kappa,\,r}}({\rm A})}{\partial
V_{\rm A}}+\frac{\partial^2 {\mathcal
I}_{_{\kappa,\,r}}({\rm B})}{\partial E_{\rm B}\,\partial
V_{\rm B}}\,S_{_{\kappa,\,r}}({\rm A})\right]\,\delta
E\,\delta V<0 \ .\\ \label{b10}
\end{eqnarray}
By using the relation
\begin{eqnarray}
\nonumber \frac{\,\,\partial^2{\mathcal
I}_{_{\kappa,\,r}}}{\partial X\,\partial Y}
&=&\frac{\kappa^2\,S_{_{\kappa,\,r}}-r\,{\mathcal
I}_{_{\kappa,\,r}}}{{\mathcal
I}_{_{\kappa,\,r}}-r\,S_{_{\kappa,\,r}}}\,\frac{\,\,\partial^2
S_{_{\kappa,\,r}}}{\partial X\,\partial Y}\\
\nonumber&+&(\kappa^2-r^2)\,\frac{{\mathcal
I}_{_{\kappa,\,r}}^2-\kappa^2\,S_{_{\kappa,\,r}}^2}{({\mathcal
I}_{_{\kappa,\,r}}-r\,S_{_{\kappa,\,r}})^3}\,\frac{\,\partial
S_{_{\kappa,\,r}}}{\partial X}\,\frac{\,\partial
S_{_{\kappa,\,r}}}{\partial Y} \ ,\\
\end{eqnarray}
Eq. (\ref{b10}) becomes
\begin{eqnarray}
\nonumber & &{1\over2}\,\left({\mathcal
I}_{_{\kappa,\,r}}-r\,S_{_{\kappa,\,r}}\right)\Bigg|_{{\rm
A}\cup{\rm B}}\\
\nonumber&\times&\Bigg[{{\mathcal S}_{_{\rm EE}}\,(\delta
E)^2+2\,{\mathcal S}_{_{\rm EV}}\,\delta E\,\delta
V+{\mathcal S}_{_{\rm VV}}\,(\delta V)^2\over{\mathcal
I}_{_{\kappa,\,r}}-r\,S_{_{\kappa,\,r}}}\,\Bigg|_{\rm A}\\
\nonumber &+&\hspace{2mm}{{\mathcal S}_{_{\rm EE}}\,(\delta
E)^2+2\,{\mathcal S}_{_{\rm EV}}\,\delta E\,\delta
V+{\mathcal S}_{_{\rm VV}}\,(\delta V)^2\over{\mathcal
I}_{_{\kappa,\,r}}-r\,S_{_{\kappa,\,r}}}\,\Bigg|_{\rm
B}\Bigg]<0 \ ,\\ \label{b12}
\end{eqnarray}
where we have defined
\begin{equation}
{\mathcal S}_{_{\rm XY}}=\frac{\,\,\partial^2
S_{_{\kappa,\,r}}}{\partial X\,\partial
Y}-\frac{(\kappa^2+r^2)\,S_{_{\kappa,\,r}}-2\,r\,{\mathcal
I}_{_{\kappa,\,r}}}{({\mathcal
I}_{_{\kappa,\,r}}-r\,S_{_{\kappa,\,r}})^2}\,\frac{\,\partial
S_{_{\kappa,\,r}}}{\partial X}\,\frac{\,\partial
S_{_{\kappa,\,r}}}{\partial Y} \ ,
\end{equation}
and taking into account Eq. (\ref{b14}), from Eq. (\ref{b12}) it
follows that the inequality
\begin{eqnarray}
{\mathcal S}_{_{\rm EE}}\,(\delta E)^2+2\,{\mathcal S}_{_{\rm
EV}}\,\delta E\,\delta V+{\mathcal S}_{_{\rm VV}}\,(\delta V)^2<0
\ ,\label{b16}
\end{eqnarray}
must holds for both systems A and B.\\
In Eq. (\ref{b16}), (by posing $\delta V=0$ and $\delta E=0$,
respectively), it follows
\begin{equation}
{\cal S}_{_{\rm EE}}<0 \ ,\hspace{10mm} {\mathcal S}_{\rm
VV}<0 \ ,\label{prima}
\end{equation}
and, multiplying Eq. (\ref{b16}) by ${\mathcal S}_{_{\rm EE}}$, we
obtain
\begin{equation}
\left({\mathcal S}_{_{\rm EE}}\,\delta E+{\mathcal S}_{_{\rm
EV}}\,\delta V\right)^2+\left({\mathcal S}_{_{\rm EE}}\,{\mathcal
S}_{_{\rm VV}}-{\mathcal S}_{_{\rm EV}}^2\right)\,(\delta V)^2>0 \
,\label{b17}
\end{equation}
which implies
\begin{equation}
{\mathcal S}_{_{\rm EE}}\,{\mathcal S}_{_{\rm VV}}-{\mathcal
S}_{_{\rm EV}}^2>0 \ .\label{b19}
\end{equation}
Eqs. (\ref{prima}) and (\ref{b19}) are the thermodynamic stability
conditions.\\ In particular, Eqs. (\ref{prima}) can be written in
\begin{equation}
\frac{\,\,\partial^2 S_{_{\kappa,\,r}}}{\partial X^2}<{\cal
A}_{_{\kappa,\,r}}\,\left(\frac{\,\partial
S_{_{\kappa,\,r}}}{\partial X}\right)^2 \ ,\label{b18}
\end{equation}
with
\begin{equation}
{\cal
A}_{_{\kappa,\,r}}=\frac{(\kappa^2+r^2)\,S_{_{\kappa,\,r}}-2\,r\,{\mathcal
I}_{_{\kappa,\,r}}}{({\mathcal
I}_{_{\kappa,\,r}}-r\,S_{_{\kappa,\,r}})^2} \ ,
\end{equation}
whilst from Eq. (\ref{b19}) we obtain
\begin{eqnarray}
\nonumber & &\frac{\,\,\partial^2
S_{_{\kappa,\,r}}}{\partial E^2}\,\frac{\,\,\partial^2
S_{_{\kappa,\,r}}}{\partial V^2}-\left(\frac{\,\,\partial^2
S_{_{\kappa,\,r}}}{\partial E\,\partial V}\right)^2>{\cal
A}_{_{\kappa,\,r}}\,{\cal B}_{_{\kappa,\,r}} \
,\\\label{b20}
\end{eqnarray}
where
\begin{eqnarray}
\nonumber {\mathcal
B}_{_{\kappa,\,r}}\!\!&=&\!\!\left(\frac{\partial^2
S_{_{\kappa,\,r}}}{\partial
E^2}\right)^{\!\!-1}\!\!\left\{\left(\frac{\partial^2
S_{_{\kappa,\,r}}}{\partial E^2}\frac{\partial
S_{_{\kappa,\,r}}}{\partial V}-\frac{\partial^2
S_{_{\kappa,\,r}}}{\partial E\,\partial V}\frac{\partial
S_{_{\kappa,\,r}}}{\partial
E}\right)^{\!\!2}\right.\\
\nonumber &+&\!\!\left.\left(\frac{\partial
S_{_{\kappa,\,r}}}{\partial
E}\right)^{\!\!2}\left[\frac{\partial^2S_{_{\kappa,\,r}}}{\partial
E^2}\,\frac{\partial^2S_{_{\kappa,\,r}}}{\partial V^2}-
\left(\frac{\partial^2 S_{_{\kappa,\,r}}}{\partial
E\,\partial V}\right)^2\right]\right\} \ ,\label{bb2}
\end{eqnarray}
and, according to Eq. (\ref{st1})-(\ref{st2}), it
follows ${\mathcal B}_{_{\kappa,\,r}}<0$.\\


\vfill\eject

\begin{thebibliography}{99}


\bibitem{Jaynes} E.T. Jaynes, ``Papers on Probability, Statistics and Statistical
Physics'', R.D. Rosenkrantz editors, (Kluwer Academic Publishers,
Dortrecht, 1989).


\bibitem{Callen} H.B. Callen, {\em Thermodynamics and an
Introduction to Thermostatistics}, (Wiley, New York, 1985).


\bibitem{Chandler} D. Chandler, {\em Introduction to Modern
Statistical Mechanics}, (Oxford University Press, Oxford, 1987).


\bibitem{Zanette} D.H. Zanette, and P.A. Alemany, Phys. Rev.
Lett. {\bf 75}, 366 (1995).


\bibitem{Compte} A. Compte, and D. Jou, J. Phys. A {\bf29}, 4321
(1996).


\bibitem{Beck} C. Beck, Physica A {\bf277}, 115 (2000).


\bibitem{Makino} J. Makino, Astrophys. J. {\bf141}, 796 (1996).


\bibitem{Uys} H. Uys, H.G. Miller, and F.C. Khanna, Phys. Lett. A
{\bf289}, 264 (2001).


\bibitem{Tanatar} B. Tanatar, Phys. Rev. E {\bf65}, 046105
(2002).


\bibitem{Rossani}A. Rossani, A.M. Scarfone, Physica A {\bf282},
212 (2000).


\bibitem{Upadhyaya} A. Upadhyaya, J.-P. Rieu, J.A. Glazier, Y.
Sawada, Physica A  {\bf293}, 549 (2001).


\bibitem{Abe} S. Abe, ``Nonextensive
Statistical Mechanics and its Applications'', Y. Okamoto editors,
(Springer 2001).


\bibitem{Kaniadakis0} Special issue of Physica A {\bf 305},
Nos. 1/2 (2002), edited by G. Kaniadakis, M. Lissia, and A.
Rapisarda.


\bibitem{Kaniadakis00} Special issue of Physica A {\bf
340},
Nos. 1/3 (2004), edited by G. Kaniadakis, and M. Lissia.


\bibitem{Naudts1}J. Naudts, Physica A {\bf340}, 32 (2004).


\bibitem{Naudts2} J. Naudts, Physica A {\bf316}, 323 (2002).


\bibitem{Ramshaw} J.D. Ramshaw, Phys. Lett. A {\bf198}, 119
(1995).


\bibitem{Wada1} T. Wada, Phys. Lett. A {\bf297}, 334 (2002).


\bibitem{Wada2} T. Wada, Continuum Mech. Thermodyn. {\bf16}, 263 (2004).


\bibitem{Wada3} T. Wada, Physica A {\bf340}, 126 (2004).


\bibitem{Taneja1} B.D. Sharma, and I.J. Taneja, Metrika {\bf 22},
205 (1975).


\bibitem{Taneja2}  B.D. Sharma, and D.P. Mittal, J. Math. Sci.
{\bf 10}, 28 (1975).


\bibitem{Mittal} D.P. Mittal, Metrika {\bf22}, 35 (1975).


\bibitem{Borges} E.P. Borges, and I. Roditi, Phys. Lett. A {\bf
246}, 399 (1998).


\bibitem{Tsallis} C. Tsallis, J. Stat. Phys. {\bf 52}, 479 (1988).


\bibitem{Tsallisweb} For a full bibliography see
http://tsallis.cat.cbpf.br/biblio.htm


\bibitem{Abe1} S. Abe, Phys. Lett. A {\bf 224}, 326 (1997).


\bibitem{Abe7} S. Abe, Phys. Lett. A {\bf 244}, 229 (1998).


\bibitem{Kaniadakis3} G. Kaniadakis, Physica A {\bf296}, 405
(2001).


\bibitem{Kaniadakis2} G. Kaniadakis, Phys. Rev. E {\bf66},
056125 (2002).


\bibitem{Scarfone1} G. Kaniadakis, M.
Lissia, and A.M. Scarfone, Phys. Rev. E (2005), {\bf71},
046128 (2005).


\bibitem{Scarfone2} G. Kaniadakis, M. Lissia, and A.M. Scarfone,
Physica A {\bf 340}, 41 (2004).


\bibitem{Hancock} H. Hancock, {\em Theory of maxima and minima},
(Dover, New York, 1960).


\bibitem{Abe2} S. Abe, Phys. Rev.
E {\bf63}, 061105 (2001).


\bibitem{Abe3} S. Abe, Physica A {\bf305}, 62 (2002).


\bibitem{Abe5} S. Abe, Physica A {\bf300}, 417 (2001).


\bibitem{Abe6} S. Abe, S. Mart\'{\i}nez, F. Pennini, and A.
Plastino, Phys. Lett. A {\bf281}, 126 (2001).


\bibitem{Wang} Q.A. Wang, L. Nivanen, and A. Le M\'{e}haut\'{e},
Eur. Phys. Lett. {\bf65}, 606 (2004).


\bibitem{Toral} R. Toral, Physica A {\bf317}, 209 (2003).

\bibitem{Velazquez} L. Velazquez, and F. Guzman, Phys. Rev. E
{\bf65}, 046134 (2002).

\bibitem{Borges1} E.P. Borges, Physica A {\bf340}, 95
(2004).

\bibitem{Wang1} L. Nivanen, A. Le M\'{e}haut\'{e}, and Q.A.
Wang, Rep. Math. Phys. {\bf52}, 437 (2003).


\bibitem{Abe8} S. Abe, Phys. Rev.
E {\bf66}, 046134 (2002).

\bibitem{Lesche} B. Lesche, J. Stat. Phys. {\bf27}, 419
(1982); Phys. Rev. E {\bf70}, 017102 (2004).


\bibitem{Zheng} C. Tsallis, A.R. Plastino, and W.-M. Zheng,
Chaos Soliton Fractals {\bf8}, 885 (1997).


\bibitem{Gelmann} M. Gell-Mann, and C. Tsallis, eds.
``Nonextensive Entropy - Interdisciplinary applications'',
(Oxford University Press, New York, 2004).


\bibitem{Guerberoff} G.R. Guerberoff, and G.A. Raggio, Phys. Lett.
A {\bf214}, 313 (1996).


\bibitem{Scarfone4} S. Abe, G. Kaniadakis, and A.M.
Scarfone, J. Phys. A: Math. Gen. {\bf37}, 10513 (2004).


\bibitem{Lissia} R. Tonelli, G. Mezzorani, F. Meloni, M.
Lissia, and M. Curaddu, ``Entropy production and Pesin
identity at the onset of chaos'', arXiv:cond -mat/0412730,
(submitted).


\bibitem{Scarfone3} G. Kaniadakis, and A.M. Scarfone,
Physica A {\bf305}, 69 (2002).


\bibitem{Tsallisrays} C. Tsallis, J.C. Anjos, and E.P.
Borges, Phys. Lett. A {\bf310}, 372 (2003).


\bibitem{Scarfonefrac} M. Cravero, G. Iabichino, G.
Kaniadakis, E. Miraldi, and A.M. Scarfone, Physica A
{\bf340}, 410 (2004).


\bibitem{Scarfone5} G. Kaniadakis, and A.M. Scarfone,
Physica A {\bf340}, 102 (2004).


\bibitem{Chakrabarti} R. Chakrabarti, and R. Jagannathan, J. Phys.
A {\bf24}, L711 (1991).


\end{thebibliography}
\end{document}